\documentclass[a4paper,11pt]{article}

\usepackage{jcappub} 

\usepackage[T1]{fontenc} 

\title{Cosmological constraints on the interacting viscous dark matter and decaying vacuum energy models}

\author{Lokesh Chander}
\author{and C.P. Singh}
\affiliation{Department of Applied Mathematics, Delhi Technological University, Delhi - 110042, India}

\emailAdd{lokeshchander\_23phdam04@dtu.ac.in}
\emailAdd{cpsingh@dce.ac.in}

\abstract{We investigate an interacting dark sector scenario in which a deviation from the standard dilution law of cold dark matter is parametrized by a perturbative correction $\epsilon$, thereby inducing a dynamical vacuum component. We extend this framework by allowing dark matter to behave as an imperfect fluid with bulk viscosity. Within the Eckart formalism, we consider a power-law dependence of the bulk viscous coefficient on the dark matter energy density which yields an effective density scaling for the viscous pressure. We obtain the parameter constraints using multi-nested sampling with \texttt{Dynesty}, employing DES 5YR Type Ia Supernovae, DESI Baryon Acoustic Oscillations, Cosmic Chronometers, the local $H_0$ measurement from H0DN, and Cosmic Microwave Background distance priors. Redshift Space Distortion measurements from SDSS-IV and the DES weak-lensing constraint on $S_8$ are included to probe structure formation. We also examine the thermodynamic consistency of the cosmological scenarios and find that they satisfy the Generalized Second Law of thermodynamics. We find that bulk viscous effects can modify the late-time interaction dynamics, while their contribution is suppressed once early-Universe information is included. These results provide observational and thermodynamic constraints on departures from the standard cold dark matter evolution and a constant vacuum energy.}

\begin{document}
\maketitle
\flushbottom

\section{Introduction}\label{sec1}

The standard model of cosmology provides a compelling description of the large-scale properties of our Universe, successfully accounting for a wide range of observational phenomena including the Cosmic Microwave Background (CMB) \cite{agha20}, large-scale structure formation \cite{bern02}, and the late-time accelerated expansion \cite{ries98,perl99}. However, with the rapid influx of high-precision observational data, it has become increasingly important to explore alternative scenarios that may address some of its outstanding problems. In particular, our understanding of the dark sector remains incomplete. While the accelerated expansion of the Universe is well established observationally, the assumption of a cosmological constant as the source of dark energy is not without issues. Beyond theoretical concerns such as fine-tuning and coincidence problems \cite{wein89,carr01,peeb03,cope06}, recent observational surveys \cite{kari25,abbo24} have also hinted at possible deviations from a strictly constant dark energy component, motivating the study of dynamical alternatives.\\
\indent A wide range of extensions to the standard paradigm have been proposed in this context. Among these, dynamical dark energy models, such as quintessence \cite{tsuj13}, introduce a time-varying equation of state to account for cosmic acceleration through scalar field dynamics. Another compelling approach is the running vacuum scenario \cite{sola18a}, motivated by Quantum Field Theory (QFT) in curved spacetime which allows for an evolving vacuum energy density that usually depends on the expansion rate. The idea was originally motivated by the fact that the vacuum itself possesses energy due to zero-point fluctuations in the quantum field \cite{zeld68}. Such models provide a natural framework for addressing some of the conceptual issues associated with a strictly constant vacuum energy and have extensively been studied as a possible alternative to the concordance cosmology \cite{ozer86,free87,shap03,borg05,sola13,sola15,sola18b,zhan19,sing21,kaeo23,sola23u,brit25,chan26a,chan26b,cruz26}.\\
\indent In addition to modifying the dark energy component, alternative descriptions of the dark sector have also been explored through modifications to the properties of dark matter. In this context, non-standard dark matter models incorporating effective fluid properties have attracted attention, particularly in scenarios where deviations from the cold and pressureless assumption may influence both background dynamics and the cosmic structure growth \cite{hu98,piat11,velt11}. One particular physical aspect that has gained increasing attention in recent years is the role of bulk viscosity. In the standard cosmological framework, cosmic fluids are typically treated as perfect, i.e. they possess an isotropic pressure and have no dissipative effects such as heat conduction and viscosity. However, from the perspective of relativistic fluid dynamics, such effects are expected to arise in realistic systems \cite{ecka40,land87,maar96}. The inclusion of bulk viscosity, as described in Eckart theory, introduces an effective negative pressure that can modify the expansion history and alter the evolution of density perturbations. This provides a physically motivated mechanism for extending the standard description of cosmic fluids and has motivated extensive investigations of viscous cosmological models in a variety of contexts, including inflationary cosmology, late-time cosmic acceleration, unified dark sector scenarios, structure formation, modified gravity theories, and thermodynamic evolution of the Universe \cite{barro86,zimd96,brev05,fab06,sing07,sing08,sing18,sing18a,simr23,avel09,norm16,barb17,most18,avel25}. Actually, within the framework of Eckart formalism, the bulk viscous perturbations can propagate at infinite speed. The Eckart theory is the first-order non-causal viscous theory which was further revised by Landau and Lifshitz \cite{land87}. Later on, Israel and Stewart \cite{isra79} proposed a full second-order causal viscous theory which addresses the causality problem. Despite the causality issues in Eckart formalism, it is frequently employed due to its simplicity. It is also to be noted that Israel-Stewart theory reduces to the Eckart theory when the relaxation time is zero in the late-time evolution of the Universe. \\
\indent In this work, we construct a generalized cosmological scenario incorporating the aforementioned elements. While interacting dark sector models and their viscous extensions have been widely investigated, with several recent studies exploring their cosmological implications \cite{sara21,cruz23,sing24,nand24,khat25a}, here we adopt a phenomenological construction in which deviations from the standard dark matter dilution law determine the evolution of the vacuum component. A particularly relevant example is the recent work \cite{khat25}, where the authors have considered a viscous interacting dark sector within the running vacuum model (RVM) framework, with the vacuum evolution specified through a Renormalization Group (RG) motivated form. In contrast, we consider a perturbative modification to the standard dilution law of dark matter, parametrized by $\epsilon$. This departure from the canonical cold dark matter (CDM) scaling induces an energy exchange within the dark sector, from which the evolution of the non-constant vacuum component is subsequently derived rather than prescribed a priori. For the viscous extension, we consider a power-law dependence of the bulk viscous coefficient on the dark matter energy density. This framework therefore allows us to investigate an alternative realization of an interacting dark sector, in which deviations from the standard dark matter dilution law and bulk viscous effects jointly modify the cosmological dynamics.

\indent We further investigate the linear perturbation dynamics of the model in the synchronous gauge at subhorizon scales \cite{ma95}. Due to the dynamical nature of vacuum and the dark sector interaction, the behavior of density perturbations differs from the standard scenario. The presence of viscous effects alters the effective pressure and can influence the growth rate of structures. We incorporate the Large Scale Structure (LSS) data including the redshift space distortion (RSD) measurements and a weak lensing constraint to probe the growth history \cite{perc10}. The vacuum component is assumed to be non-clustering, consistent with its interpretation as a smooth background contribution in running vacuum scenarios \cite{basi14,gome17}.\\
\indent For the observational analysis, we employ the \texttt{dynesty} nested sampling algorithm \cite{spea20}, which directly computes the Bayesian evidence together with posterior parameter distributions using the nested sampling framework originally developed by Skilling \cite{skil04,skil06}. The analysis utilizes the multi-ellipsoidal bounding and random slice sampling implementation available within the \texttt{dynesty} package \cite{buch17}. The datasets used include Type Ia Supernovae from the Dark Energy Survey (DES 5YR) \cite{abbo24}, Baryon Acoustic Oscillation measurements from the Dark Energy Spectroscopic Instrument (DESI) Data Release II \cite{kari25}, and Cosmic Chronometer data \cite{more20}, which together probe the background expansion history. We additionally include the $H_0$ constraint obtained from the $H_0$ Distance Network (H0DN) collaboration \cite{stef26}. For the perturbation analysis, we utilize the LSS data including RSD measurements from the Sloan Digital Sky Survey (SDSS-IV) Data Release 17 \cite{alam21} and the DES estimate of the clustering parameter $S_8$ \cite{abbo25} to constrain structure formation. We statistically analyse the model using selection criteria including the Akaike Information Criterion (AIC), Bayesian Information Criterion (BIC), and Bayes Factor \cite{aka73,sch78,kass95}. \\
\indent We also analyse the model from a thermodynamic point of view, as bulk viscosity modifies the nature of the dark matter component and introduces dissipative effects. The dark sector interaction induces an energy exchange which also modifies background thermodynamics. We review the second law of thermodynamics by adopting the Bekenstein--Hawking formalism \cite{beke73,hawk75} for the horizon entropy and check whether the Generalized Second Law (GSL) remains valid. \\
\indent The structure of this paper is as follows. In Sec. \ref{sec2}, we introduce the theoretical framework of the model, providing the necessary background on the dark sector interaction and bulk viscosity. In Sec. \ref{sec3}, we derive the governing equations and obtain analytical expressions for the Hubble expansion rate and other cosmographic parameters. The interaction dynamics and perturbation analysis, including the evolution of the matter density contrast, are presented in Sec. \ref{sec4}. In Sec. \ref{sec5}, we describe the observational datasets and the statistical methodology employed for parameter estimation. The resulting Bayesian posterior constraints and statistical model comparison are discussed in Sec. \ref{sec6}. In Sec. \ref{sec7}, we examine the thermodynamic viability of the framework by confronting it with the Generalized Second Law of thermodynamics. Finally, Sec. \ref{sec8} summarizes the main results and outlines possible directions for future work.

\section{Theoretical Framework}\label{sec2}

We consider a spatially flat Friedmann-Lema\^{i}tre-Robertson-Walker (FLRW) metric given by
\begin{equation}\label{eq1}
ds^2 = -dt^2 + a^2(t)\left[dr^2 + r^2\left(d\theta^2 + \sin^2\theta\, d\phi^2\right)\right].
\end{equation}

We assume relativistic units $c=1$, with $(t,r,\theta,\phi)$ representing the temporal and spatial components in spherical coordinates. The cosmic fluid is assumed to consist of baryons, dark matter, and a vacuum component. The baryonic component follows the standard conservation equation
\begin{equation}\label{eq2}
\dot{\rho}_b + 3H\rho_b = 0 \quad \Rightarrow \quad \rho_b = \rho_{b,0} a^{-3}.
\end{equation}
Here $\rho_{b,0}$ is the present value of the baryon density. Standard cold dark matter (CDM), like baryonic matter, follows the canonical dilution law $\rho_{dm} \propto a^{-3}$. To phenomenologically describe departures from this standard evolution, we introduce a perturbative modification to the dark matter density of the form
\begin{equation}\label{eq3}
\rho_{dm} = \rho_{dm,0}a^{-3+\epsilon},
\end{equation}
where $\rho_{dm,0}$ is the value of dark matter density at present and the parameter $\epsilon$ quantifies deviations from the standard CDM dilution law. Similar modifications to the dark matter evolution have previously been considered in interacting dark sector scenarios, including more general constructions in which $\epsilon$ is allowed to depend on the scale factor, i.e. $\epsilon=\epsilon(a)$ \cite{alca05,cost10}. In this work, we consider $\epsilon$ to be a constant, representing the simplest departure from the standard CDM scenario.

A deviation from the canonical dark matter dilution law can be associated with an energy exchange between dark matter and a dynamical vacuum component. We therefore consider an interacting dark sector described by the continuity equations
\begin{equation}\label{eq4}
\begin{aligned}
\dot{\rho}_{dm} + 3H(\rho_{dm} + p_{dm}) &= Q_I, \\
\dot{\rho}_\Lambda &= - Q_I,
\end{aligned}
\end{equation}
where $\rho_\Lambda$ denotes the time-varying vacuum energy density and $Q_I$ characterizes the energy exchange between the dark matter and vacuum components. The sign convention is such that $Q_I>0$ corresponds to energy transfer from the vacuum to dark matter, while $Q_I<0$ corresponds to energy transfer in the opposite direction.

Combining the individual fluid equations in \eqref{eq4}, we obtain the conservation equation for the coupled dark sector,
\begin{equation}\label{eq5}
\dot{\rho}_{dm} + 3H(\rho_{dm} + p_{dm}) = -\dot{\rho}_\Lambda.
\end{equation}

Thus, rather than imposing an independent functional form for the evolution of the vacuum energy density, we determine $\rho_\Lambda(a)$ through the dark sector conservation equation. Once $\rho_\Lambda(a)$ is obtained, the corresponding interaction term $Q_I$ can subsequently be reconstructed from \eqref{eq4}.\\

\noindent In the standard cosmological model, dark matter is treated as a pressureless perfect fluid, characterized by the equilibrium pressure $p_{dm}=0$ and the absence of dissipative effects. In this work, we relax this assumption and model dark matter as a non-perfect fluid endowed with bulk viscosity. This leads to an effective pressure that can modify both the background evolution and the growth of cosmic structures.

In the Eckart formalism \cite{ecka40}, the energy-momentum tensor of a viscous fluid is given by
\begin{equation}\label{eq6}
T_{\mu\nu} = (\rho + \bar p)u_\mu u_\nu + \bar p\, g_{\mu\nu},
\end{equation}
where $u^\mu$ is the four-velocity of the fluid and $\bar p = p + \Pi$ denotes the effective pressure consisting of the equilibrium pressure $p$ and the bulk viscous contribution $\Pi$. The corresponding continuity equation follows from the conservation law $\nabla_\nu T^{\mu\nu}=0$.

\indent Thus, the effective pressure for the viscous dark matter component can be written as
\begin{equation}\label{eq7}
\bar{p}_{dm} = p_{dm} + \Pi = p_{dm} - 3 \zeta H,
\end{equation}
where $\Pi=-3\zeta H$ denotes the bulk viscous pressure in the Eckart formalism, $H$ is the Hubble expansion rate, and $\zeta$ is the bulk viscous coefficient. The latter characterizes the departure of the fluid from local thermodynamic equilibrium and is required to satisfy $\zeta>0$ in accordance with the thermodynamic principles. For dark matter, we retain the intrinsic pressureless condition $p_{dm}=0$, such that all effective pressure contributions arise purely from bulk viscous effects.\\

\indent The bulk viscous coefficient may depend on the properties of the cosmic fluid, including its energy density and the expansion rate. Generalized parametrizations incorporating such dependencies have been considered in the literature, for example $\zeta \sim H^{1-2s}\rho^s$ \cite{gab23}. Other phenomenological forms in which the bulk viscous coefficient scales directly with the Hubble expansion rate, as well as more general algebraic extensions, have also been explored. In this work, we adopt a phenomenological parametrization involving both the dark matter energy density and the Hubble expansion rate, motivated by similar constructions previously considered for viscous dark-sector components \cite{eeh17}.
We consider the bulk viscous coefficient in the form
\begin{equation}\label{eq8}
\zeta \equiv \xi_0 \frac{\rho_{dm}^{1/2}}{3H},
\end{equation}
where $\xi_0>0$ is a constant parameter. During the matter-dominated era, the Friedmann equation implies $H^2\approx {\rho_{dm}}$ in reduced Planck units, such that the bulk viscous coefficient approaches an effectively constant value. Therefore, $\xi_0$ carries the same dimensions as that of the bulk viscous coefficient $\zeta$, i.e., of inverse time.

\section{Analytical Solutions}\label{sec3}

The Friedmann equation for the system and its time derivative are given by
\begin{align}\label{e9,10}
\frac{3H^2}{8\pi G} &= \rho_b + \rho_{dm} + \rho_\Lambda, \\
\frac{6H\dot{H}}{8\pi G} &= \dot{\rho}_b + \dot{\rho}_{dm} + \dot{\rho}_\Lambda.
\end{align}

In Eq. \eqref{eq5} instead of $p_{dm}$, we will consider the effective pressure for the dark matter component $\bar{p}_{dm}$ given in Eq. \eqref{eq7}. Substituting $\zeta$ from Eq. \eqref{eq8}, and rewriting in terms of the scale factor $a$, \eqref{eq5} becomes
\begin{equation}\label{eq12}
\frac{d\rho_{dm}}{da} + \frac{3}{a} \left( \rho_{dm} - \xi_0 \rho_{dm}^{1/2} \right)
= -\frac{d\rho_\Lambda}{da}.
\end{equation}

\noindent Using the modified dilution law for dark matter in \eqref{eq3}, we arrive at the following differential equation:
\begin{align}\label{eq13}
\frac{d\rho_\Lambda}{da} = 3 \xi_0 \rho_{dm,0}^{1/2} a^{\frac{-5+\epsilon}{2}} - \epsilon \rho_{dm,0} a^{-4+\epsilon}
\end{align}

\noindent To obtain the analytic expression for vacuum energy density, we must integrate the above within the proper limits:
\begin{align}\label{eq14}
\rho_\Lambda(a) &= \rho_{\Lambda,0} + \int_{1}^{a} (3 \xi_0 \rho_{dm,0}^{1/2} a^{\frac{-5+\epsilon}{2}} - \epsilon \rho_{dm,0} a^{-4+\epsilon}) da,
\end{align}
where $\rho_{\Lambda,0}$ is the value of the vacuum energy density at present ($a=1$). After solving and rearranging the terms, we obtain the expression as:
\begin{align}\label{eq15}
\rho_\Lambda(a) &= \rho_{\Lambda,0} + \frac{\epsilon \rho_{dm,0}}{3 - \epsilon} \left(a^{-3+\epsilon} - 1 \right) - \frac{6\xi_0 \rho_{dm,0}^{1/2}}{(3-\epsilon)} \left(a^{\frac{-3+\epsilon}{2}} - 1 \right).
\end{align}

\noindent Using the Friedmann equation, we express the Hubble parameter as
\begin{equation}\label{eq16}
H^2 = \frac{8\pi G}{3} \left[\rho_{b,0} a^{-3} + \rho_{dm,0} a^{-3+\epsilon} + \rho_\Lambda(a) \right].
\end{equation}

Substituting $\rho_\Lambda(a)$ from Eq. \eqref{eq15}, and dividing by the critical density $\rho_{c,0} = 3 H_0^2 / 8 \pi G$, we obtain
\begin{align}\label{eq17}
H^2(a) = H_0^2 \Bigg[
&\, \Omega_{b} a^{-3}
+ \Omega_{dm} \left(
\frac{3a^{-3+\epsilon} - \epsilon}{3-\epsilon}
\right) + \Omega_{\zeta} \frac{3(a^{\frac{-3+\epsilon}{2}}- 1)}{3-\epsilon}
+ \Omega_{\Lambda}
\Bigg],
\end{align}
where $\Omega_{i} = \rho_{i,0}/\rho_{c,0}$ and in particular we define the dimensionless parameter for the bulk viscous contribution $\Omega_\zeta$ as:
\begin{equation}\label{eq18}
\Omega_{\zeta} \equiv -\frac{2\xi_0 \rho_{dm,0}^{1/2}}{\rho_{c,0}}.
\end{equation}
Given that $\xi_0>0$, the above parameter is required to be negative to ensure thermodynamic consistency. The boundary condition gives $\Omega_b + \Omega_{dm} + \Omega_{\Lambda} = 1$. Moreover, we can write the Hubble parameter concisely using the rescaled density parameters as:
\begin{align}\label{eq19}
H(a) = H_0 (
&\Omega_{b} a^{-3}
+ \tilde{\Omega_{dm}} a^{-3+\epsilon}
+ \tilde{\Omega_{\zeta}} a^{\frac{-3+\epsilon}{2}}
+ \tilde{\Omega_{\Lambda}})^{1/2}.
\end{align}
where
\begin{equation}
\begin{aligned}\label{eq20}
  \tilde{\Omega_{dm}} \equiv \frac{3}{3-\epsilon} \Omega_{dm} , \quad \tilde{\Omega_{\zeta}} \equiv \frac{3}{3-\epsilon} \Omega_\zeta , \\ \quad
  \tilde{\Omega_{\Lambda}} \equiv \Omega_{\Lambda} - \frac{\epsilon}{3-\epsilon} \Omega_{dm} - \frac{3}{3-\epsilon} \Omega_\zeta.
\end{aligned}
\end{equation}
These parameters capture the effect of bulk viscosity on the interacting dark sector induced by the perturbative factor $\epsilon$.  We also have $\Omega_b + \tilde{\Omega_{dm}} + \tilde{\Omega_{\zeta}} + \tilde{\Omega_{\Lambda}} = 1$. \\

\noindent In addition to the Hubble parameter, we also discuss the following cosmological parameters which are important for probing the expansion history and dynamics of the universe.

\textit{Deceleration Parameter:}
The deceleration parameter quantifies the acceleration of the Universe and is defined as
\begin{equation}\label{eq21}
q(a) = -1 - \frac{\dot{H}}{H^2} = -1 - \frac{a}{H}\frac{dH}{da}.
\end{equation}
A negative value of $q$ indicates accelerated expansion. Substituting $H$ from Eq. \eqref{eq19}, and using the relations in \eqref{eq20}, we obtain the deceleration parameter for the model as:
\begin{equation}\label{eq22}
q(a) = -1 + \frac{3}{2 E^2} \left( {\Omega_ba^{-3} + \Omega_{dm}a^{-3+\epsilon} +\frac{1}{2} \Omega_{\zeta} a^{\frac{-3+\epsilon}{2}}}\right),
\end{equation}
where $E$ is the dimensionless Hubble parameter given as $E(a)=H(a)/H_0$.

\textit{Transition Redshift:}
The transition redshift $z_t$ denotes the redshift at which the expansion of the Universe switches from deceleration to acceleration. This quantity is obtained by imposing the condition that the deceleration parameter vanishes. i.e.
\begin{equation}\label{eq23}
q(a = a_t) = 0.
\end{equation}
Here $a_t$ is the value of the scale factor at which transition occurs and the respective redshift can be obtained via the relation $a_t = (1 + z_t)^{-1}$.\\

\textit{Effective Equation of State:}
The effective equation-of-state (EoS) parameter is defined in terms of the Hubble parameter as
\begin{equation}\label{eq24}
w(a) = -1 - \frac{2}{3}\frac{\dot{H}}{H^2}
= -1 - \frac{2a}{3H}\frac{dH}{da}.
\end{equation}
This provides a measure of the overall cosmic fluid behavior. The expression for the effective EoS parameter for the model is:
\begin{equation}\label{eq25}
w(a) = -1 + \frac{1}{E^2} \left( {\Omega_ba^{-3} + \Omega_{dm}a^{-3+\epsilon} +\frac{1}{2} \Omega_{\zeta} a^{\frac{-3+\epsilon}{2}}}\right),
\end{equation}

\textit{Jerk Parameter:}
The jerk parameter is a higher-order kinematical quantity useful for distinguishing cosmological models, defined as
\begin{equation} \label{eq26}
j(a) = \frac{\dddot{a}}{aH^3}
= q(2q+1) + a\frac{dq}{da}.
\end{equation}
A closed form analytic expression of $j(a)$ for our model can be obtained by subsituting Eq. \eqref{eq22} into the above. For the $\Lambda$CDM model, $j=1$ at all times.\\

\textit{Age of the Universe:}
The age of the Universe, $t_0$, represents the time elapsed since the Big Bang to the present epoch. It can be obtained by integrating the inverse Hubble expansion rate over the cosmic history,

\begin{equation}\label{eq27}
t_0 = \int_0^1 \frac{da}{aH(a)}.
\end{equation}

Using the expression for H(a) given in Eq.~\eqref{eq19}, the age of the Universe for the model can be written as

\begin{equation}\label{eq28}
t_0=
\frac{1}{H_0}
\int_0^1
\frac{da}
{a\left(
\Omega_{b} a^{-3}
+\tilde{\Omega}_{dm} a^{-3+\epsilon}
+\tilde{\Omega}_{\zeta} a^{\frac{-3+\epsilon}{2}}
+\tilde{\Omega}_{\Lambda}
\right)^{1/2}}.
\end{equation}

\section{Linear Perturbations with Interaction Dynamics}\label{sec4}

In this section we study the interaction dynamics and the evolution of perturbations in the dark matter sector. For the vacuum component, we assume there is no clustering as the decaying vacuum follows the equation of state $p_{\Lambda} = - \rho_{\Lambda}$. We assume density perturbations with vacuum dynamics at subhorizon scale in the synchronous gauge.

\subsection{Dark Sector Interaction}

The interaction between the dark matter and dynamical vacuum components is characterized by the term $Q_I$ appearing in the continuity equations \eqref{eq4}. Once the modified dark matter evolution and the viscous contribution are specified, the interaction term can be obtained from the background conservation equations. We find
\begin{equation}\label{eq29}
Q_I = \epsilon_{*}\rho_{dm}H,
\end{equation}
where
\begin{equation}\label{eq30}
\epsilon_{*} =
\epsilon + \frac{3}{2}\frac{\Omega_{\zeta}}{\Omega_{dm}}
a^{\frac{3-\epsilon}{2}}.
\end{equation}

The second term in $\epsilon_{*}$ arises from the bulk viscous contribution, making the effective interaction parameter redshift dependent. During the matter-dominated era, where $a\ll1$, the viscous contribution is strongly suppressed, such that $\epsilon_{*}\approx\epsilon$. Consequently, the interaction term approaches
\begin{equation}
Q_I \simeq \epsilon\rho_{dm}H,
\end{equation}
at early times, while the viscous contribution becomes increasingly relevant at later epochs.

Since both $\rho_{dm}$ and $H$ are positive, the sign of $\epsilon_{*}$ determines the direction of energy transfer. With the convention adopted in Eq.~\eqref{eq4}, $\epsilon_{*}>0$ corresponds to energy transfer from the vacuum sector to dark matter, whereas $\epsilon_{*}<0$ implies energy transfer from dark matter to the vacuum sector \cite{wang16}.

\subsection{Growth of DM Perturbations}

\indent We consider the evolution of dark matter perturbations in the sub-horizon regime, $k^2 \gg a^2H^2$. In the present treatment, we neglect pressure-gradient contributions to the dark matter perturbations. This choice is motivated by the known difficulties associated with treating perturbations in the Eckart formalism, which can lead to an effective negative sound speed and consequently to an unstable pressure-gradient contribution. We therefore assume that the dark matter perturbations cluster in a manner analogous to cold dark matter on the scales considered here. The interaction with the dynamical vacuum component introduces additional terms in the perturbation evolution, which modify the effective friction and source terms governing the growth of matter perturbations. The corresponding perturbation equations were derived by G\'{o}mez-Valent and Sol\`a Peracaula within the running vacuum model (RVM) framework \cite{gome18}. Since our scenario likewise contains an interacting dark matter-vacuum sector, we adopt the same perturbative formalism.
\\

\noindent The modified growth equation can be written as
\begin{equation}\label{eq31}
\delta_{dm}'' + \mathcal{A}(a)\,\delta_{dm}' + \mathcal{B}(a)\,\delta_{dm} = 0,
\end{equation}
where the friction term $\mathcal{A}(a)$ and the source term $\mathcal{B}(a)$ are given by
\begin{align}\label{eq32,33}
\mathcal{A}(a) &= \frac{3}{a} + \frac{H'}{H} + \frac{\Psi}{aH}, \\
\mathcal{B}(a) &= - \frac{3}{2}\frac{\Omega_{dm}(a)}{a^2}
+ \frac{2\Psi}{a^2 H} + \frac{\Psi'}{aH}.
\end{align}

Here prime denotes differentiation with respect to the scale factor `$a$' and the interaction is encoded in the term $\Psi$ as,
\begin{equation}\label{eq34}
\Psi = -\frac{\dot{\rho}_\Lambda}{\rho_{dm}} = \frac{Q_I}{\rho_{dm}},
\end{equation}

\noindent Using Eq. \eqref{eq29}, we can write $\Psi$ as
\begin{equation}\label{eq35}
\Psi(a) = H(a)\left[
\epsilon + \frac{3}{2}\frac{\Omega_{\zeta}}{\Omega_{dm}}\,
a^{\frac{3-\epsilon}{2}}
\right].
\end{equation}


\indent As evident from Eq.\eqref{eq31}, the interaction manifests itself through an extra friction term and an altered source term, both of which influence the growth of matter perturbations compared to the standard $\Lambda$CDM framework. The growth equation is solved numerically from an initial scale factor $a_i=10^{-3}$ up to the present epoch $(a=1)$, using the initial conditions given in \cite{silv21}.\\

\noindent The dark matter density parameter $\Omega_{dm}(a)$ can be written as:
\begin{equation}\label{eq36}
    \Omega_{dm}(a) = \frac{\Omega_{dm}(a=1) a^{-3+\epsilon}}{E^2 (a)}
\end{equation}

\indent We note that on subhorizon scales, the baryon and dark matter perturbations evolve similarly, such that $\delta_b \simeq \delta_{dm}$. Since dark matter constitutes the dominant fraction of the total matter density, the total matter contrast $\delta_m$ can be approximated by the dark matter density contrast. Consequently, the normalized growth function may be expressed in terms of $\delta_{dm}$ as
\begin{equation}\label{eq37}
    D_m(a) = \frac{\delta_{dm} (a)}{\delta_{dm}(a=1)}.
\end{equation}

\noindent The linear growth rate is given by
\begin{equation}\label{eq38}
    f(a) = \frac{d \ln D_m(a)}{d \ln a}.
\end{equation}

\indent In interacting scenarios, the observable growth rate inferred from redshift-space distortions is modified due to the energy transfer \cite{yang18}. The corrected growth rate in terms of the redshift $z$ is given by
\begin{equation}\label{eq39}
    f_{\rm RSD}(z) = f(z) - \frac{Q_I}{H\rho_{dm}}.
\end{equation}

The parameter $\sigma_8(z)$, which quantifies the amplitude of matter fluctuations smoothed over spheres of radius $8h^{-1} \mathrm{Mpc}$, evolves according to
\begin{equation}\label{eq40}
\sigma_8(z)=\frac{\delta_m(z)}{\delta_m(0)}\sigma_{8,0}.
\end{equation}
Here, $\sigma_{8,0} \equiv \sigma_8(0)$ is the present-day value of $\sigma_8$. In our analysis, we denote it simply by $\sigma_8$ and constrain it as a free parameter of the model with the LSS data.

The quantity directly constrained by redshift-space distortion measurements is $f\sigma_8(z)$, defined as the product of the linear growth rate and the fluctuation amplitude,
\begin{equation}\label{eq41}
f\sigma_8(z)=f_{\rm RSD}(z)\sigma_8(z).
\end{equation}

Since both $f_{\rm RSD}$ and $\delta_m$ are determined by the underlying cosmological dynamics, the resulting $f\sigma_8$ predictions naturally encode the combined effects of the background expansion history and the interaction sector on the growth of large-scale structure.

\section{Data and Methodology}\label{sec5}

This section presents the observational datasets and methodology used to constrain the model parameters. The analysis is carried out within a Bayesian framework, where the likelihood is assumed to follow a Gaussian distribution,
\begin{equation}\label{eq42}
\mathcal{L} \propto \exp\left(-\frac{\chi^2}{2}\right).
\end{equation}
The chi-squared statistic is constructed from the differences between the theoretical predictions and the corresponding observational data.

We consider three models: the standard $\Lambda$CDM model, the interacting decaying vacuum model (IDVM), and its viscous extension (VIDVM). These models are subsequently confronted with a range of cosmological observations described below.

\subsection{Late-Time Background Probes}
To constrain the background expansion history of the Universe, we employ a combination of late-time geometric probes sensitive to cosmological distances and the Hubble expansion rate. These datasets provide complementary constraints across a broad redshift range and are widely used in cosmology analyses.\\
\textbf{Type Ia Supernovae (SNIa)}: We use the five-year Type Ia Supernovae sample obtained from the Dark Energy Survey (DES) \cite{abbo24}, consisting of 1635 photometrically classified supernovae in the redshift range $0.10<z<1.13$. To provide a robust low-redshift anchor, this sample is combined with an external compilation of 194 spectroscopically confirmed supernovae spanning $0.025<z<0.10$. The full dataset therefore comprises 1829 distinct supernovae. The supernova likelihood construction follows the standard distance modulus formalism and analytic marginalization procedure commonly adopted in cosmological analyses \cite{trip98,conl11,scol18}.\\
\textbf{Baryon Acoustic Oscillations (BAO)}: We use the 2025 BAO measurements from the Dark Energy Spectroscopic Instrument (DESI) DRII release \cite{kari25} which includes means, standard deviations, and the full covariance matrix of the observables. Within the redshift range $0.295 \leq z \leq 2.330$, there are 7 effective redshifts at which measurements are made: 1 for $D_V/r_d$, and 6 each for $D_M/r_d$ and $D_H/r_d$, accounting for a total of 13 measurements. The BAO likelihood analysis follows the standard methodology based on the comoving angular diameter distance, Hubble distance, and sound horizon scale commonly employed in cosmological parameter estimation \cite{eise05,alam17}.\\
\textbf{Cosmic Chronometers (CC)}: We use 32 Cosmic Chronometer (CC) measurements of the Hubble parameter $H(z)$ obtained from the differential age technique \cite{jime02} applied to passively evolving galaxies over the redshift range $0.07 \leq z \leq 1.965$. The analysis incorporates the full covariance matrix together with the associated statistical and systematic uncertainties of the dataset \cite{more12,more16,more18,more20}.\\
\textbf{Local Hubble Measurement (LHM)}: Recent distance ladder methods have estimated the local value of the Hubble constant $H_0$ to be higher than Planck estimates, indicating a faster expansion rate. To reconcile the tension between these two different observational methods we include the Baseline measurement of $H_0$, reported as $ 73.50 \pm 0.81 $ km/s/Mpc by the $H_0$ Distance Network (H0DN) Collaboration \cite{stef26}.\\

\subsection{Large Scale Structure (LSS)}

In addition to the background observations, we include large scale structure probes sensitive to the growth of matter perturbations and structure formation, allowing us to test deviations from the standard $\Lambda$CDM growth history induced by the viscous and interacting dark sector modifications.\\
\textbf{Redshift Space Distortions (RSD)}: The perturbation evolution is modified due to the interaction dynamics in the dark sector. Since dark matter dominates the late-time matter distribution, we apply the modified perturbation equations to the total matter sector. We use 6 RSD measurements from SDSS-IV \cite{alam21} and adopt the methodology described in Section \ref{sec4} for the analysis.\\
\textbf{Weak Lensing Constraint}: Weak gravitational lensing provides an additional probe of structure growth through the clustering parameter $S_8$, which depends on both the matter density and the amplitude of matter fluctuations. We incorporate the DES weak-lensing measurement together with the RSD data to obtain tighter constraints on the growth sector, adopting the observed value $S_8 = 0.776 \pm 0.017$ \cite{abbo25}.

\subsection{High-Redshift Probes}

In the present work, we investigate cosmological scenarios involving a variable vacuum energy density together with interacting dark matter modified through bulk viscous effects. These deviations from the standard $\Lambda$CDM scenario can modify the high-redshift expansion history and consequently alter quantities such as the sound horizon scale at drag epoch $r_d$, the acoustic scale $l_A$, and the shift parameter $R$. \\
\textbf{Cosmic Microwave Background (CMB) Priors}: These priors provide geometric constraints on the early-universe evolution through quantities such as the shift parameter $R$, the acoustic scale $l_A$, and the baryon density parameter $\omega_b$, enabling us to test whether the interacting and viscous sectors remain compatible with the observed pre-recombination cosmology \cite{chen19,wang07}.\\
\textbf{Early-Universe Calibration}: In analyses where CMB priors are not included, we adopt the standard Big Bang Nucleosynthesis (BBN)-consistent baryon density value $\omega_b = \Omega_b h^2 = 0.0224$ \cite{cybu16} and fix the sound horizon scale to $r_d = 147.1\,{\rm Mpc}$, consistent with the Planck 2018 results \cite{agha20}.\\

\subsection{Datasets}

We use three different combinations of observational datasets in order to constrain the cosmological parameters and investigate the impact of the bulk viscosity and interacting dark sector components at both the background and perturbative levels.

\begin{itemize}

\item \textbf{BASE}: This dataset combination consists of the late-time background probes including Type Ia Supernovae (SNIa), Baryon Acoustic Oscillations (BAO), Cosmic Chronometers (CC), and the local Hubble parameter measurement ($H_0$). The total number of effective data points in this combination is $N=1875$. The corresponding total chi-square function is defined as
\begin{equation}\label{eq43}
\chi^2_{\rm tot} = \chi^2_{\rm SNIa} + \chi^2_{\rm BAO} + \chi^2_{\rm CC} + \chi^2_{\rm {H_0}}.
\end{equation}

\item \textbf{BASE + LSS}: In order to probe the growth of cosmic structures, we extend the BASE dataset by incorporating late-time LSS observations through RSD and the DES weak lensing constraint for $S_8$. This allows us to constrain the perturbative growth parameters $\sigma_8$ and $S_8$. The total number of effective data points for this dataset combination becomes $N=1882$. The corresponding chi-square function is given by
\begin{equation}\label{eq44}
\chi^2_{\rm tot} = \chi^2_{\rm BASE} + \chi^2_{f\sigma_8} + \chi^2_{S_8},
\end{equation}
where $ \chi^2_{f\sigma_8} $ and $ \chi^2_{S_8} $ are the contributions from the LSS data.

\item \textbf{BASE + LSS + CMB}: Finally, we supplement the previous dataset combination with the CMB distance priors, which provide an effective early-time geometric anchor on the cosmological evolution. The inclusion of the CMB priors enables tighter constraints on the matter density and expansion history parameters. The total number of effective data points in this full dataset combination is $N=1885$. The corresponding total chi-square function is expressed as
\begin{equation}\label{eq45}
\chi^2_{\rm tot} = \chi^2_{\rm BASE} + \chi^2_{\rm LSS} + \chi^2_{\rm CMB}.
\end{equation}

\end{itemize}
The parameter space is explored using the nested sampling algorithm \texttt{dynesty} \cite{spea20}, which efficiently computes posterior distributions and Bayesian evidence. Flat priors were assumed for all free cosmological parameters during the analysis. Specifically, the Hubble parameter was varied within the range $H_0 \in [60,80]$, the dark matter density parameter within $\Omega_{dm} \in [0.1,0.4]$, the bulk viscous parameter within the small negative range $\Omega_\zeta \in [-0.03,0]$, the interaction parameter within $\epsilon \in [-0.3,0.3]$, the clustering amplitude within $\sigma_8 \in [0.6,1.1]$ and the baryon density parameter $\omega_b \in [0.01, 0.03]$. The best-fit values correspond to the minimum of $\chi^2$, while confidence intervals are derived from the marginalized posterior distributions.


\begin{table}[ht]
\centering
\caption{Cosmological constraints for the BASE dataset combination (SNIa + BAO + CC + Local $H_0$). The quoted values correspond to the median and $1\sigma$ credible intervals obtained from the posterior distributions.}
\label{t1}
\begin{tabular}{lccc}
\hline
\hline
Parameter & $\Lambda$CDM & IDVM & VIDVM \\
\hline

$H_0$ [km/s/Mpc] &
$69.39^{+0.40}_{-0.43}$ &
$73.24^{+0.87}_{-0.79}$ &
$69.00^{+0.48}_{-0.46}$ \\

$\Omega_{dm}$ &
$0.248^{+0.006}_{-0.006}$ &
$0.241^{+0.007}_{-0.008}$ &
$0.267^{+0.014}_{-0.015}$ \\

$\epsilon$ &
--- &
$-0.144^{+0.002}_{-0.002}$ &
$0.093^{+0.062}_{-0.068}$ \\

$\Omega_{\zeta}$ &
--- &
--- &
$-0.0029^{+0.0022}_{-0.0046}$ \\

\hline

$\Omega_{\Lambda}$ &
$0.705^{+0.007}_{-0.007}$ &
$0.718^{+0.007}_{-0.008}$ &
$0.686^{+0.014}_{-0.015}$ \\

$q_0$ &
$-0.558^{+0.010}_{-0.010}$ &
$-0.576^{+0.011}_{-0.011}$ &
$-0.535^{+0.023}_{-0.021}$ \\

$z_t$ &
$0.685^{+0.019}_{-0.019}$ &
$0.650^{+0.018}_{-0.018}$ &
$0.686^{+0.018}_{-0.020}$ \\

$w_0$ &
$-0.705^{+0.007}_{-0.007}$ &
$-0.717^{+0.008}_{-0.007}$ &
$-0.690^{+0.016}_{-0.014}$ \\

$j_0$ &
$1.000^{+0.000}_{-0.000}$ &
$1.052^{+0.001}_{-0.001}$ &
$0.968^{+0.027}_{-0.028}$ \\

$t_0 \, [\mathrm{Gyr}]$ &
$13.65^{+0.04}_{-0.04}$ &
$12.71^{+0.17}_{-0.16}$ &
$13.75^{+0.08}_{-0.08}$ \\

\hline
\hline
\end{tabular}
\end{table}

\begin{table}[ht]
\centering
\caption{Cosmological constraints for the BASE+LSS dataset combination (SNIa + BAO + CC + Local $H_0$ + RSD + $S_8$). The quoted values correspond to the median and $1\sigma$ credible intervals obtained from the posterior distributions.}
\label{t2}
\begin{tabular}{lccc}
\hline
\hline
Parameter & $\Lambda$CDM & IDVM & VIDVM \\
\hline

$H_0$ [km/s/Mpc] &
$69.52^{+0.40}_{-0.41}$ &
$73.15^{+0.73}_{-0.77}$ &
$69.48^{+0.40}_{-0.41}$ \\

$\Omega_{dm}$ &
$0.246^{+0.006}_{-0.006}$ &
$0.248^{+0.007}_{-0.007}$ &
$0.243^{+0.007}_{-0.008}$ \\

$\sigma_8$ &
$0.802^{+0.018}_{-0.017}$ &
$0.763^{+0.018}_{-0.017}$ &
$0.796^{+0.017}_{-0.018}$ \\

$\epsilon$ &
--- &
$-0.142^{+0.002}_{-0.002}$ &
$-0.029^{+0.022}_{-0.022}$ \\

$\Omega_{\zeta}$ &
--- &
--- &
$-0.0032^{+0.0024}_{-0.0048}$ \\

\hline

$\Omega_{\Lambda}$ &
$0.708^{+0.007}_{-0.007}$ &
$0.711^{+0.007}_{-0.008}$ &
$0.710^{+0.008}_{-0.007}$ \\

$S_8$ &
$0.792^{+0.015}_{-0.014}$ &
$0.749^{+0.016}_{-0.014}$ &
$0.782^{+0.016}_{-0.017}$ \\

$q_0$ &
$-0.561^{+0.010}_{-0.010}$ &
$-0.566^{+0.011}_{-0.010}$ &
$-0.571^{+0.011}_{-0.011}$ \\

$z_t$ &
$0.691^{+0.018}_{-0.019}$ &
$0.634^{+0.017}_{-0.018}$ &
$0.690^{+0.017}_{-0.018}$ \\

$w_0$ &
$-0.708^{+0.007}_{-0.007}$ &
$-0.711^{+0.008}_{-0.007}$ &
$-0.714^{+0.008}_{-0.007}$ \\

$j_0$ &
$1.000^{+0.000}_{-0.000}$ &
$1.053^{+0.001}_{-0.001}$ &
$1.015^{+0.008}_{-0.008}$ \\

$t_0 \, [\mathrm{Gyr}]$ &
$13.66^{+0.04}_{-0.04}$ &
$12.65^{+0.16}_{-0.15}$ &
$13.62^{+0.05}_{-0.05}$ \\

\hline
\hline
\end{tabular}
\end{table}

\begin{table*}[ht]
\centering
\caption{Cosmological constraints for the BASE+LSS+CMB dataset combination (SNIa + BAO + CC + Local $H_0$ + RSD + $S_8$ + CMB distance priors). The quoted values correspond to the median and $1\sigma$ credible intervals obtained from the posterior distributions.}
\label{t3}
\begin{tabular}{lccc}
\hline
\hline
Parameter & $\Lambda$CDM & IDVM & VIDVM \\
\hline

$H_0$ [km/s/Mpc] &
$69.09^{+0.26}_{-0.27}$ &
$69.30^{+0.43}_{-0.48}$ &
$69.33^{+0.46}_{-0.48}$ \\

$\Omega_{dm}$ &
$0.249^{+0.003}_{-0.003}$ &
$0.249^{+0.003}_{-0.003}$ &
$0.250^{+0.003}_{-0.004}$ \\

$\omega_b$ &
$0.02265^{+0.00011}_{-0.00012}$ &
$0.02268^{+0.00014}_{-0.00014}$ &
$0.02267^{+0.00013}_{-0.00014}$ \\

$\sigma_8$ &
$0.797^{+0.015}_{-0.017}$ &
$0.795^{+0.016}_{-0.016}$ &
$0.790^{+0.018}_{-0.018}$ \\

$\epsilon$ &
--- &
$-0.002^{+0.003}_{-0.003}$ &
$-0.002^{+0.003}_{-0.003}$ \\

$\Omega_{\zeta}$ &
--- &
--- &
$-0.0087^{+0.0073}_{-0.0077}$ \\

\hline

$\Omega_{\Lambda}$ &
$0.704^{+0.003}_{-0.003}$ &
$0.703^{+0.003}_{-0.003}$ &
$0.702^{+0.004}_{-0.004}$ \\

$r_d$ [Mpc] &
$146.834^{+0.176}_{-0.190}$ &
$146.123^{+1.124}_{-1.195}$ &
$145.874^{+1.170}_{-1.166}$ \\

$S_8$ &
$0.792^{+0.015}_{-0.016}$ &
$0.791^{+0.016}_{-0.014}$ &
$0.787^{+0.016}_{-0.017}$ \\

$q_0$ &
$-0.556^{+0.005}_{-0.005}$ &
$-0.555^{+0.005}_{-0.005}$ &
$-0.564^{+0.009}_{-0.008}$ \\

$z_t$ &
$0.681^{+0.009}_{-0.008}$ &
$0.679^{+0.009}_{-0.009}$ &
$0.686^{+0.011}_{-0.010}$ \\

$w_0$ &
$-0.704^{+0.003}_{-0.003}$ &
$-0.703^{+0.003}_{-0.003}$ &
$-0.709^{+0.006}_{-0.006}$ \\

$j_0$ &
$1.000^{+0.000}_{-0.000}$ &
$1.001^{+0.001}_{-0.001}$ &
$1.011^{+0.009}_{-0.008}$ \\

$t_0 \, [\mathrm{Gyr}]$ &
$13.69^{+0.02}_{-0.02}$ &
$13.64^{+0.09}_{-0.09}$ &
$13.61^{+0.09}_{-0.09}$ \\

\hline
\hline
\end{tabular}
\end{table*}

\begin{figure*}
\centering
\includegraphics[width=0.65\textwidth]{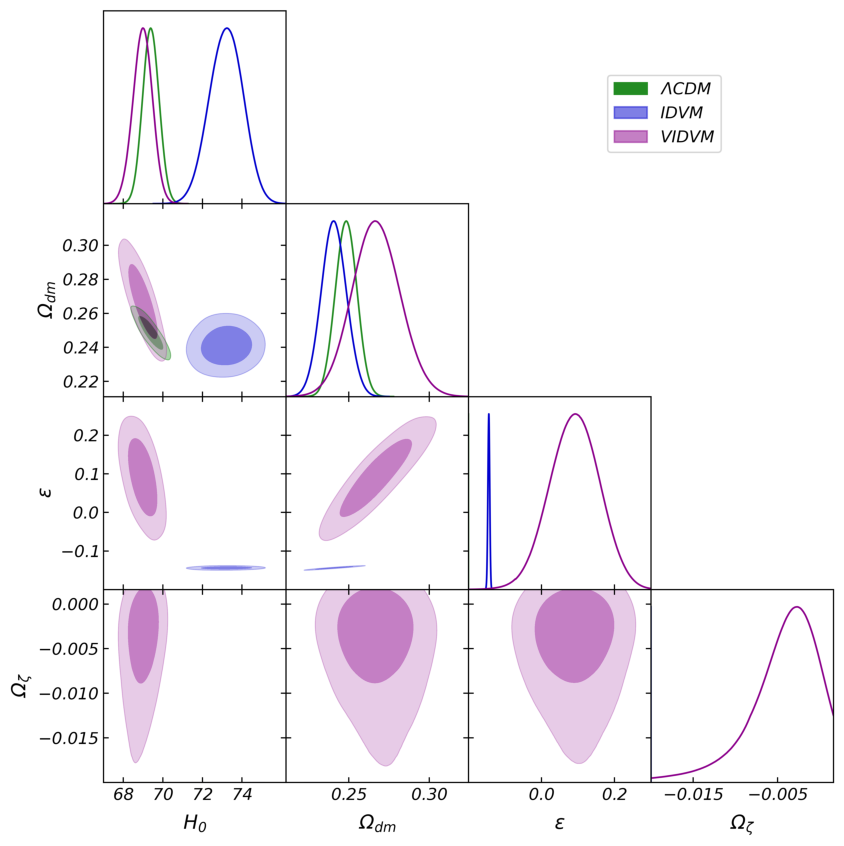}
\caption{Joint posterior distributions and marginalized likelihoods for the BASE dataset combination (SNIa + BAO + CC + local $H_0$) obtained for the three cosmological scenarios: $\Lambda$CDM, IDVM, and VIDVM. The interacting dark vacuum model exhibits a significant shift toward higher values of $H_0$ together with a strongly constrained negative interaction parameter $\epsilon$, while the inclusion of the viscous sector partially restores the contours toward the standard $\Lambda$CDM region.}\label{fig1}
\end{figure*}

\begin{figure*}
\centering
\includegraphics[width=0.75\textwidth]{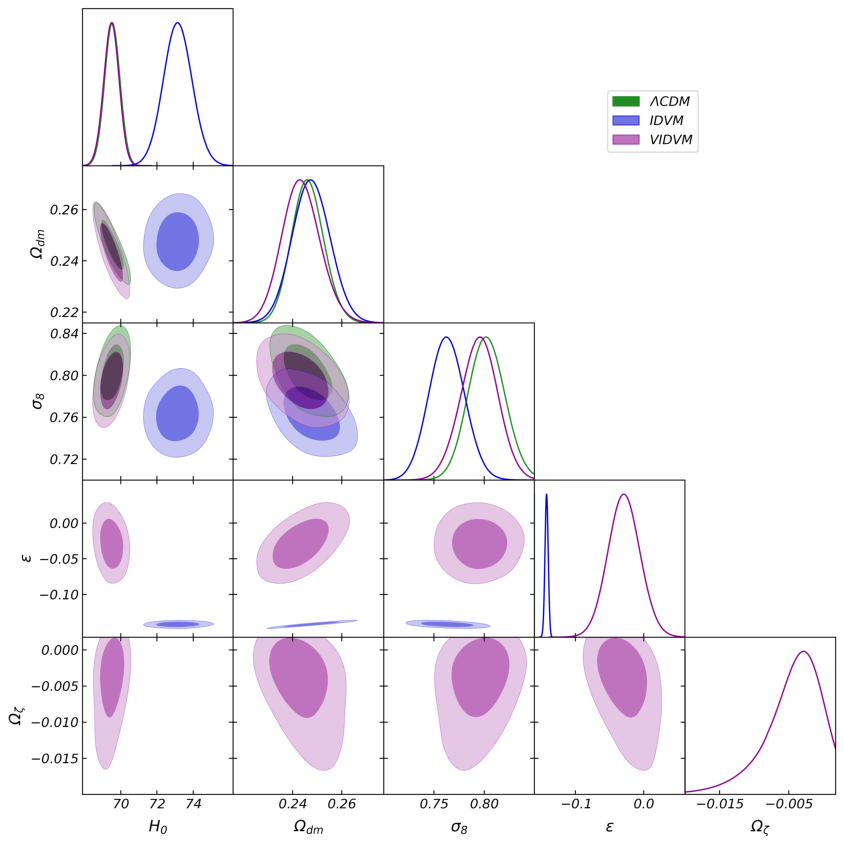}
\caption{Joint posterior distributions and marginalized likelihoods for the BASE+LSS dataset combination (SNIa + BAO + CC + local $H_0$ + RSD + $S_8$). The inclusion of late-time structure growth measurements does not affect the strongly interacting branch in IDVM. The VIDVM scenario yields intermediate behaviour with interaction getting suppressed to a small negative value, partially alleviating the suppression in $\sigma_8$ induced by the pure interaction sector.}\label{fig2}
\end{figure*}

\begin{figure*}
\centering
\includegraphics[width=0.85\textwidth]{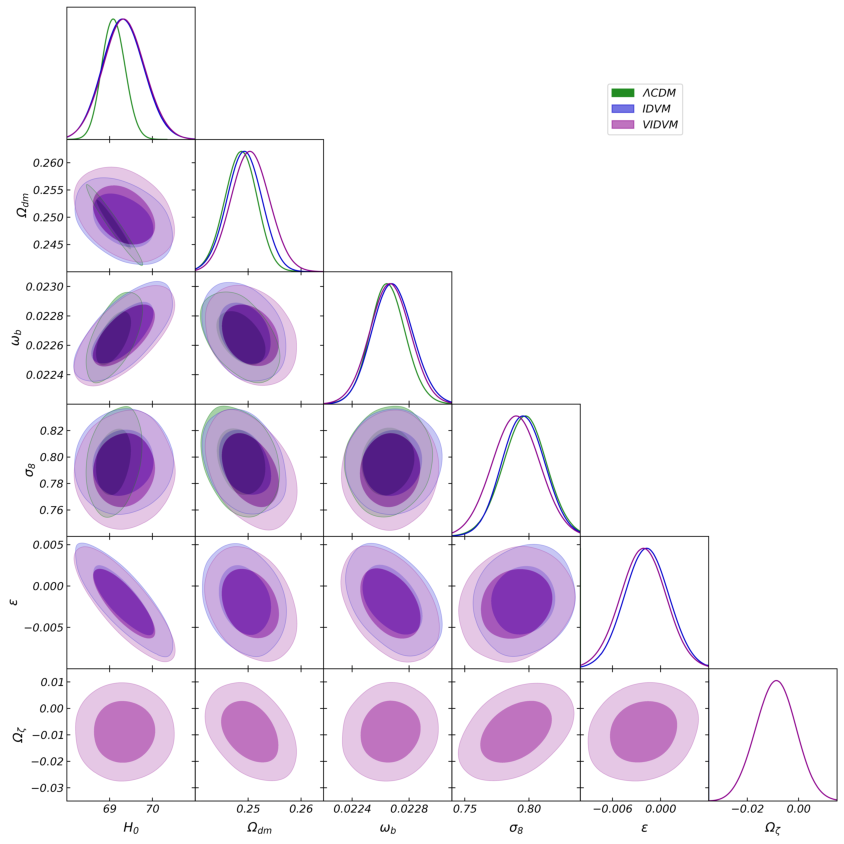}
\caption{Joint posterior distributions and marginalized likelihoods for the full BASE+LSS+CMB dataset combination (SNIa + BAO + CC + local $H_0$ + RSD + $S_8$ + CMB distance priors). The inclusion of CMB distance priors significantly tightens the parameter constraints and drives both extended scenarios toward the $\Lambda$CDM limit. Consequently, the contours for IDVM and VIDVM largely overlap with those of the standard cosmological model. While the interaction parameter $\epsilon$ is strongly suppressed, the bulk viscous parameter $\Omega_{\zeta}$ remains weakly non-zero.}\label{fig3}
\end{figure*}

\section{Results and Discussion}\label{sec6}

First, we discuss the cosmological parameter constraints obtained from the nested sampling analysis for the different dataset combinations. We then examine the statistical performance of the models through a comparison of the $\chi^2$ contributions, together with model selection criteria such as the reduced chi-square, AIC, BIC, and Bayesian evidence.

\subsection{Parameter Constraints}

\begin{figure*}
\centering
\includegraphics[width=0.85\textwidth]{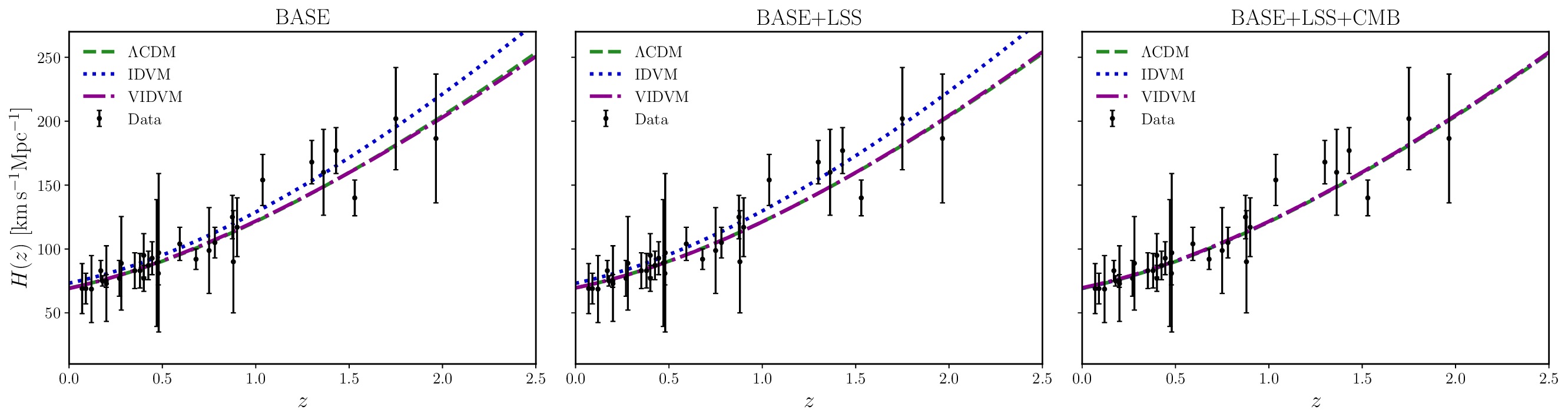}
\caption{Evolution of the Hubble expansion rate $H(z)$ for the $\Lambda$CDM, IDVM, and VIDVM scenarios using the BASE, BASE+LSS, and BASE+LSS+CMB dataset combinations. The cosmic chronometer (CC32) measurements are shown for comparison.}\label{fig4}
\end{figure*}

\begin{figure*}
\centering
\includegraphics[width=0.85\textwidth]{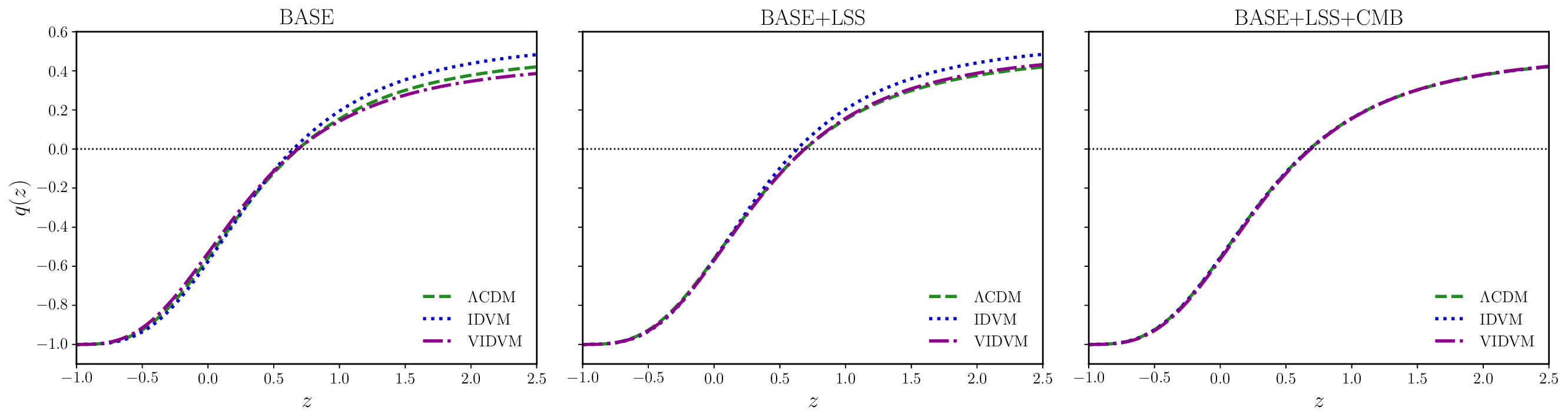}
\caption{Evolution of the deceleration parameter $q(z)$ for the $\Lambda$CDM, IDVM, and VIDVM models for the three dataset combinations considered in this work. All models exhibit the standard transition from an early decelerating phase to the present accelerated expansion epoch.}\label{fig5}
\end{figure*}

\begin{figure*}
\centering
\includegraphics[width=0.85\textwidth]{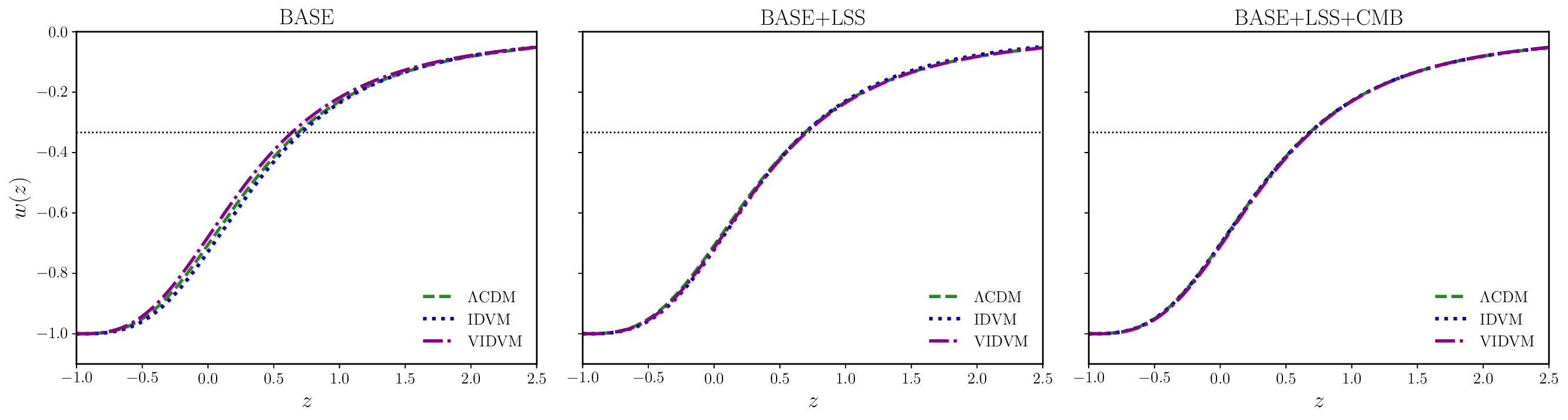}
\caption{Evolution of the effective equation of state parameter $w(z)$ for the $\Lambda$CDM, IDVM, and VIDVM scenarios. The trajectories become increasingly similar with the inclusion of successive datasets.}\label{fig6}
\end{figure*}

\begin{figure*}
\centering
\includegraphics[width=0.85\textwidth]{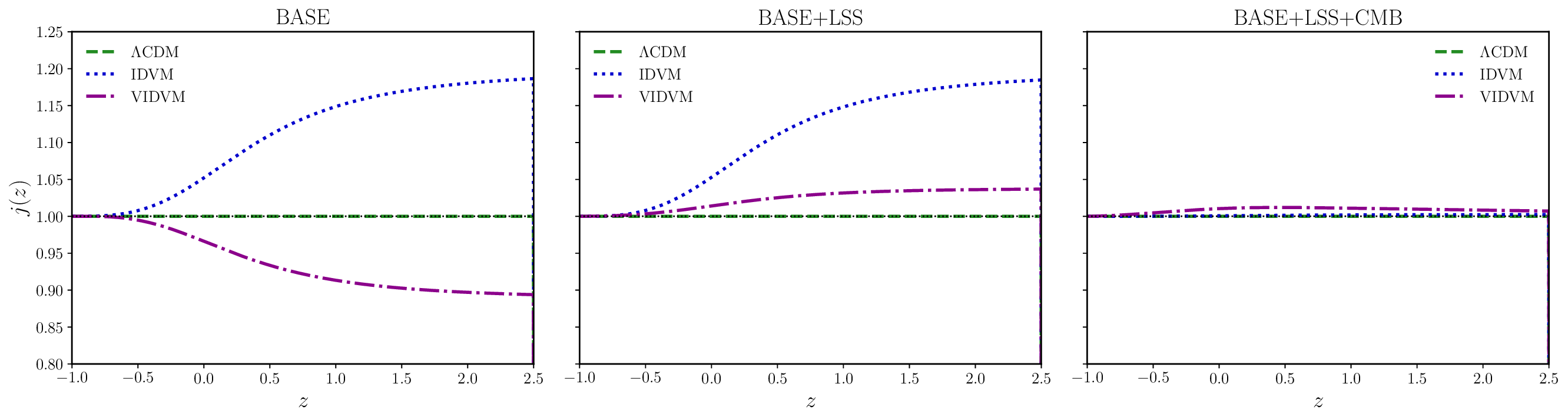}
\caption{Evolution of the jerk parameter $j(z)$ for the $\Lambda$CDM, IDVM, and VIDVM cosmologies using the best-fit parameters obtained from the BASE, BASE+LSS, and BASE+LSS+CMB dataset combinations. The standard $\Lambda$CDM model predicts a constant value $j=1$ at all redshifts, shown by the horizontal green curve.}
\label{fig7}
\end{figure*}

The cosmological constraints obtained from the three observational dataset combinations are summarized in Tables~\ref{t1}, \ref{t2}, and \ref{t3} for the BASE, BASE+LSS, and BASE+LSS+CMB datasets, respectively. The contour plots of the model's parameters are shown in Figs.\ref{fig1}-\ref{fig3}. For the BASE dataset, the inferred values of the Hubble constant are $H_0 = 69.39^{+0.40}_{-0.43}$, $73.24^{+0.87}_{-0.79}$, and $69.00^{+0.48}_{-0.46}\ {\rm km\,s^{-1}\,Mpc^{-1}}$ for the $\Lambda$CDM, IDVM, and VIDVM models, respectively. A notable feature is that the interacting vacuum model yields a significantly higher value of $H_0$, bringing it considerably closer to the local Hubble measurement. The inclusion of viscous dark matter substantially reduces the preferred value of $H_0$, thereby increasing its tension with the local determination.
For the BASE+LSS dataset, a similar trend is observed, with $H_0 = 69.52^{+0.40}_{-0.41}$, $73.15^{+0.73}_{-0.77}$, and $69.48^{+0.40}_{-0.41}\ {\rm km\,s^{-1}\,Mpc^{-1}}$ for the $\Lambda$CDM, IDVM, and VIDVM models, respectively. The inclusion of structure-growth information slightly lowers the preferred $H_0$ value in the IDVM scenario, although the model continues to favor a significantly higher expansion rate than the other two cosmologies. Overall, the addition of LSS data improves consistency with the growth measurements while preserving the tendency of the interacting vacuum model to alleviate the Hubble tension.
The impact of the CMB distance priors is particularly striking. Upon their inclusion, the inferred values converge toward a common region of parameter space. The $\Lambda$CDM model yields $H_0 = 69.09^{+0.26}_{-0.27}\ {\rm km\,s^{-1}\,Mpc^{-1}}$, while the IDVM and VIDVM models become nearly indistinguishable with $H_0 = 69.30^{+0.43}_{-0.48}$ and $69.33^{+0.46}_{-0.48}\ {\rm km\,s^{-1}\,Mpc^{-1}}$, respectively. This convergence highlights the strong constraining power of early-Universe information and indicates that both the interaction and viscous contributions are tightly restricted once CMB priors are taken into account.The corresponding evolution of the Hubble parameter for the best-fit estimates is shown in Fig. \ref{fig4}.\\
\indent Next we discuss the interaction dynamics of the models through the parameters $\Omega_{dm}$, $\epsilon$, and $\Omega_\zeta$. For the BASE dataset, the dark matter density is constrained as $\Omega_{dm}=0.248\pm0.006$, $0.241^{+0.007}_{-0.008}$, and $0.267^{+0.014}_{-0.015}$ for the $\Lambda$CDM, IDVM, and VIDVM models respectively. The inclusion of viscous effects therefore increases the preferred dark matter fraction relative to both $\Lambda$CDM and the pure interaction scenario.
With the addition of the LSS data, the values of $\Omega_{dm}$ become much more tightly clustered across the three models. In fact, the VIDVM model now admits the lowest value, $\Omega_{dm}=0.243^{+0.007}_{-0.008}$. This behavior can be understood from the corresponding evolution of the interaction parameter. For the BASE dataset, the viscous model prefers a positive value of $\epsilon$, which slows the dilution of dark matter and leads to a larger present-day matter density. Once the LSS measurements are included, however, the preferred interaction shifts to a negative value, reducing this effect and driving the dark matter density toward lower values.
The interaction parameter itself exhibits a remarkably stable behavior in the IDVM scenario. We obtain $\epsilon=-0.144\pm0.002$ and $\epsilon=-0.142\pm0.002$ for the BASE and BASE+LSS datasets respectively, resulting in the very sharp contours visible in Figs.~\ref{fig1} and \ref{fig2}. In contrast, the inclusion of bulk viscosity significantly weakens the constraints on the interaction strength due to the degeneracy between $\epsilon$ and $\Omega_\zeta$. The preferred values become $\epsilon=0.093^{+0.062}_{-0.068}$ and $\epsilon=-0.029\pm0.022$ for the BASE and BASE+LSS datasets, respectively. Not only are these constraints considerably broader, but the magnitude of the interaction is also reduced relative to the pure interacting case.
The bulk viscous density parameter is constrained as $\Omega_\zeta=-0.0029^{+0.0022}_{-0.0046}$ and $-0.0032^{+0.0024}_{-0.0048}$ for the BASE and BASE+LSS datasets. In both cases the parameter exhibits a long negative tail and remains only weakly constrained by late-time observations alone.
The inclusion of the CMB distance priors produces a much more consistent picture across all models. The dark matter density becomes nearly identical in the three models, while the interaction parameter is strongly suppressed to $\epsilon=-0.002\pm0.003$ in both the IDVM and VIDVM scenarios. This result indicates that a substantial dark sector interaction is strongly disfavored once early-Universe information is taken into account. More interestingly, the viscous parameter survives the inclusion of the CMB priors, with $\Omega_\zeta=-0.0087^{+0.0073}_{-0.0077}$. Although still consistent with zero at the $1\sigma$ level, this constraint is noticeably tighter than those obtained from the BASE and BASE+LSS datasets. This suggests that while a strong interaction may be ruled out by early-Universe physics, bulk viscosity could still play a subtle role in the dark sector dynamics.\\
\indent Now we discuss the kinematics of the models through the parameters $q_0$, $z_t$, $w_0$, $j_0$, and $t_0$. For the BASE dataset, the IDVM model stands out owing to its stronger accelerated expansion. Compared to $\Lambda$CDM, it prefers a more negative deceleration parameter $q_0$, while the VIDVM model exhibits a less negative value. The present-day effective equation of state parameter $w_0$ follows the same trend, with IDVM favoring a more negative value and VIDVM remaining closer to the standard cosmological picture. The transition redshift also occurs noticeably earlier in IDVM, whereas the values for $\Lambda$CDM and VIDVM remain largely comparable. A similar pattern is observed in the jerk parameter, with $j_0>1$ for IDVM and $j_0<1$ for VIDVM, indicating that the two extensions depart from the standard cosmological model in opposite directions. The corresponding age estimates are $t_0=13.65\pm0.04$, $12.71^{+0.17}_{-0.16}$, and $13.75\pm0.08\ \mathrm{Gyr}$ for $\Lambda$CDM, IDVM, and VIDVM respectively, reflecting the substantially higher expansion rate favored by the interacting vacuum model.
With the inclusion of the LSS data, the values of $q_0$ and $w_0$ move closer together across the three cosmologies, indicating that the growth measurements disfavor the more extreme expansion histories permitted by the BASE dataset alone. The transition redshift remains noticeably earlier in IDVM, while continuing to be very similar for $\Lambda$CDM and VIDVM. The jerk parameter now becomes greater than unity for both extended models, although the deviations remain modest. The age estimates decrease slightly for the extended scenarios while remaining nearly unchanged for $\Lambda$CDM. Overall, the addition of structure-growth information has a visible impact on the kinematic sector, but the qualitative trends remain similar to those observed for the BASE dataset.
The inclusion of the CMB distance priors leads to a much more uniform picture. As the interaction parameter is driven toward zero, the $\Lambda$CDM and IDVM models become effectively indistinguishable in terms of $q_0$, $z_t$, $w_0$, and $j_0$. The VIDVM model continues to exhibit mild deviations due to the residual viscous contribution, although these departures are now quite subtle. The age estimates also become considerably more consistent, with $\Lambda$CDM yielding the highest value of $t_0=13.69\pm0.02\ \mathrm{Gyr}$ and VIDVM the lowest value of $13.61\pm0.09\ \mathrm{Gyr}$.
Overall, the kinematic evolution of the three cosmological scenarios remains highly consistent. The IDVM model emerges as the clear outlier for the BASE and BASE+LSS datasets owing to its preference for a stronger accelerated expansion and a younger Universe. Once the CMB priors are included, however, the interacting vacuum model becomes virtually indistinguishable from $\Lambda$CDM, while the viscous extension retains small but non-negligible deviations in the kinematic parameters.
The plots for the deceleration parameter $q(z)$, effective EoS parameter $w(z)$ and the jerk parameter $j(z)$ are shown in Figs. \ref{fig5} - \ref{fig7} for all the models across the three datasets.

Next we come to the constraints obtained for structure growth. In our analysis, we adopted the $S_8$ estimate from DES, as the RSD data alone were insufficient to constrain the matter clustering amplitude $\sigma_8$ to physically viable values. For the BASE+LSS dataset, we obtain $\sigma_8 = 0.802^{+0.018}_{-0.017}$, $0.763^{+0.018}_{-0.017}$, and $0.796^{+0.017}_{-0.018}$ for the $\Lambda$CDM, IDVM, and VIDVM models respectively. Notably, the IDVM model yields the lowest value, indicating a significant suppression of structure growth. The derived parameter $S_8$ takes the values $0.792^{+0.015}_{-0.014}$, $0.749^{+0.016}_{-0.014}$, and $0.782^{+0.016}_{-0.017}$, again with the IDVM model exhibiting the strongest suppression and remaining in noticeable tension with the DES estimate adopted in this work.

\begin{figure*}
\centering
\includegraphics[width=0.7\textwidth]{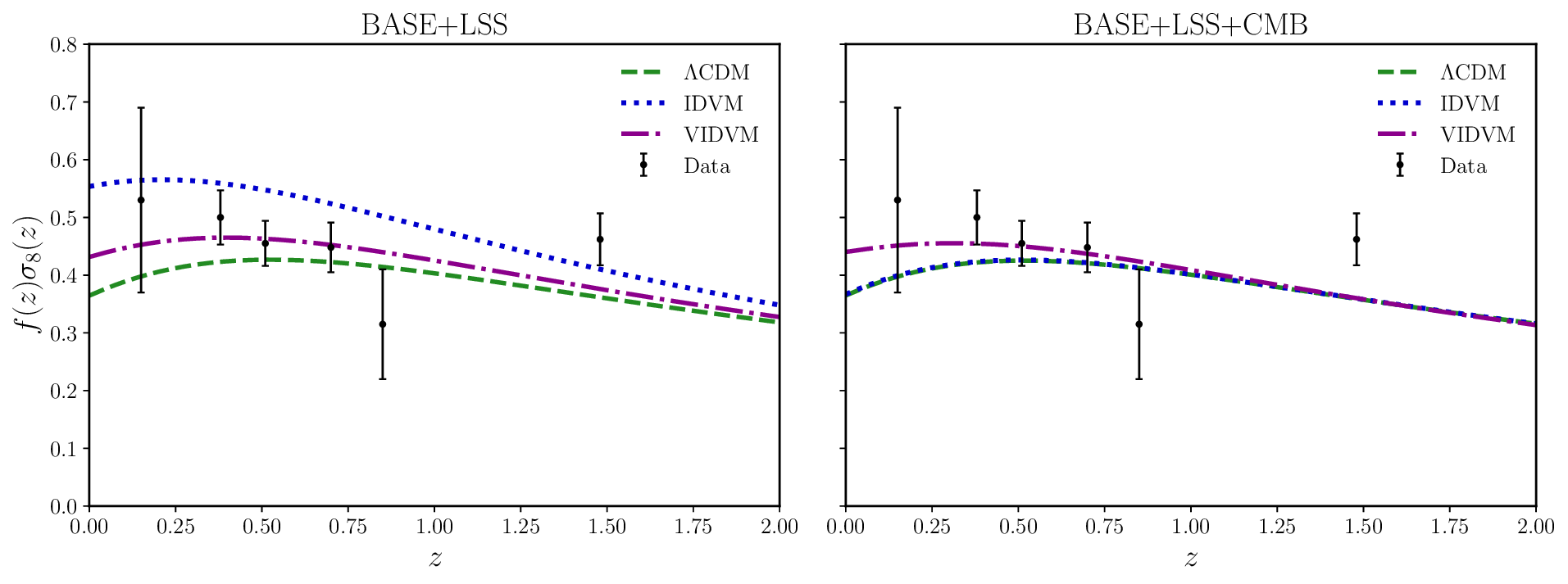}
\caption{Redshift evolution of the growth observable $f(z)\sigma_8(z)$ for the $\Lambda$CDM, IDVM, and VIDVM models compared against the SDSS-IV RSD measurements.}\label{fig8}
\end{figure*}

With the inclusion of the CMB distance priors, the clustering amplitudes become much more consistent across the three cosmological scenarios. The constraints on $\sigma_8$ tighten to $0.797^{+0.015}_{-0.017}$, $0.795^{+0.016}_{-0.016}$, and $0.790^{+0.018}_{-0.018}$, while the corresponding values of $S_8$ become $0.792^{+0.015}_{-0.016}$, $0.791^{+0.016}_{-0.014}$, and $0.787^{+0.016}_{-0.017}$. This demonstrates that once early-Universe information is included, the three models predict a very similar level of matter clustering.
The evolution of the weighted linear growth rate $f(z)\sigma_8(z)$ is shown in Fig.~\ref{fig8}. An interesting feature of the results is that although the IDVM model prefers lower values of both $\sigma_8$ and $S_8$, its $f\sigma_8$ trajectory lies above those of the other models. This behavior originates from the interaction correction appearing in the RSD growth rate, as discussed in Eq.~\eqref{eq39}. The correction term $Q_I/(H\rho_{dm})$ is precisely the effective interaction parameter $\epsilon_*$. For the BASE+LSS dataset, $\epsilon_*$ is negative and has a comparatively large magnitude in the IDVM scenario. Consequently, the interaction correction effectively enhances the observed growth rate, producing a higher $f\sigma_8$ trajectory despite the lower clustering amplitude inferred from the parameter constraints.

For the BASE+LSS+CMB dataset, the $\Lambda$CDM and IDVM trajectories become nearly indistinguishable, reflecting the strong suppression of the interaction parameter by the CMB priors. The VIDVM model, however, exhibits a noticeable departure at low redshifts. This behavior can be understood from Eq.~\eqref{eq30}. In the CMB-constrained regime, the interaction parameter $\epsilon$ becomes subdominant, while the contribution proportional to $\Omega_\zeta$ becomes increasingly relevant at late times. Through the effective interaction parameter $\epsilon_*$, this viscous contribution modifies the RSD correction and gives rise to the observed deviation in the growth history. Thus, while the CMB data drive the interaction sector toward the $\Lambda$CDM limit, the viscous component retains a measurable impact on late-time structure formation.

The CMB priors allow us to constrain the baryon density $\omega_b$ and the sound horizon scale $r_d$. We obtain nearly identical values of $\omega_b$ across all three cosmologies, while $r_d$ remains close to $146 {\rm Mpc}$ in each case. This indicates that the interacting and viscous extensions primarily affect the late-time Universe, leaving the pre-recombination physics largely unchanged.

Overall, the results illustrate how different observational datasets impact the interacting decaying vacuum scenario, both with and without bulk viscosity, relative to the standard $\Lambda$CDM cosmology. The BASE and BASE+LSS datasets favor a strong negative interaction in the IDVM model, leading to noticeable departures from the standard cosmological picture in both the background and perturbation sectors. In contrast, the inclusion of bulk viscosity weakens the preferred interaction strength, with the BASE dataset admitting a mild positive interaction and the BASE+LSS dataset favoring a much smaller negative value. The inclusion of the CMB distance priors dramatically alters this picture by driving the interaction parameter toward zero in both extended scenarios, rendering the IDVM model nearly indistinguishable from $\Lambda$CDM. Consequently, the strongest departures from the standard cosmological model arise from late-time observations alone, while the addition of early-Universe information enforces a much more restricted region of parameter space for the dark sector.

\begin{table*}[ht]
\centering
\caption{Statistical comparison of the cosmological models for different dataset combinations. The quantities $\Delta$AIC, $\Delta$BIC and $\Delta \log Z$ are computed relative to the corresponding $\Lambda$CDM model for each dataset combination.}
\label{t4}

\begin{tabular}{llcccccccc}
\hline
\hline
Dataset & Model & $\chi^2_{\rm min}$ & $\chi^2_\nu$ & AIC & $\Delta$AIC & BIC & $\Delta$BIC & $\log Z$ & $\Delta \log Z$ \\
\hline

BASE
& $\Lambda$CDM
& 1741.13 & 0.93 & 1745.14 & 0.00 & 1756.20 & 0.00 & -877.15 & 0.00 \\

& IDVM
& 1729.18 & 0.92 & 1735.19 & -9.95 & 1751.78 & -4.42 & -875.34 & +1.82 \\

& VIDVM
& 1739.28 & 0.93 & 1747.30 & +2.16 & 1769.43 & +13.23 & -879.58 & -2.43 \\

\hline

+LSS
& $\Lambda$CDM
& 1752.41 & 0.93 & 1758.43 & 0.00 & 1775.03 & 0.00 & -885.35 & 0.00 \\

& IDVM
& 1748.77 & 0.93 & 1756.79 & -1.64 & 1778.93 & +3.90 & -887.62 & -2.27 \\

& VIDVM
& 1749.31 & 0.93 & 1759.34 & +0.91 & 1787.01 & +11.98 & -888.03 & -2.68 \\

\hline

+CMB
& $\Lambda$CDM
& 1762.23 & 0.94 & 1770.25 & 0.00 & 1792.39 & 0.00 & -896.76 & 0.00 \\

& IDVM
& 1761.93 & 0.94 & 1771.96 & +1.71 & 1799.63 & +7.24 & -901.19 & -4.43 \\

& VIDVM
& 1760.66 & 0.94 & 1772.71 & +2.46 & 1805.91 & +13.52 & -903.40 & -6.64 \\

\hline
\hline
\end{tabular}
\end{table*}

\subsection{Statistical Tests}

In this section, we assess the relative statistical preference of the interacting cosmologies with respect to the standard $\Lambda$CDM model. To quantify both the goodness of fit and model complexity, we employ several statistical criteria including the minimum chi-square statistic $\chi^2_{\rm min}$, the reduced chi-square $\chi^2_\nu$, the Akaike Information Criterion (AIC), the Bayesian Information Criterion (BIC), and the Bayesian evidence $\log Z$.

The reduced chi-square is defined as
\[
\chi^2_{\nu} = \frac{\chi^2_{\rm min}}{N-k},
\]
where $N$ denotes the total number of data points and $k$ represents the number of free model parameters. Values of $\chi^2_{\nu}$ close to unity indicate statistically acceptable fits to the observational data.

To compare models with different numbers of free parameters, we additionally compute the Akaike Information Criterion (AIC) \cite{aka73} given by
\[
\mathrm{AIC} = \chi^2_{\rm min} + 2k,
\]
which penalizes models with larger parameter spaces while favoring improved goodness-of-fit. Lower AIC values correspond to statistically preferred models.

We also evaluate the Bayesian Information Criterion \cite{sch78} defined as
\[
\mathrm{BIC} = \chi^2_{\rm min} + k \ln N,
\]
which imposes a stronger penalty on models with additional parameters, particularly for large observational datasets.

Finally, Bayesian model comparison is performed using the Bayesian evidence $\log Z$ \cite{kass95} calculated through the \texttt{Dynesty} nested sampling algorithm \cite{spea20}. In this framework, models with larger evidence values are statistically favored. The relative preference between competing models may be quantified through the Bayes factor,
\[
\Delta \log Z = \log Z_i - \log Z_j,
\]
where positive values indicate preference for model $i$ relative to model $j$. We use the Jeffrey's scale \cite{trot08} to interpret the strength of evidence, with $|\Delta \log Z| < 1$ indicating inconclusive evidence, $1 \leq |\Delta \log Z| < 2.5$ corresponding to weak evidence, $2.5 \leq |\Delta \log Z| < 5$ representing moderate evidence, and $|\Delta \log Z| \geq 5$ indicating strong evidence in favor of the preferred model. The resulting statistical estimators for all models and dataset combinations are summarized in Table \ref{t4}.

\indent For the BASE dataset combination, the IDVM scenario yields the lowest minimum chi-square value with $\chi^2_{\rm min}=1729.18$, compared to $\chi^2_{\rm min}=1741.13$ for $\Lambda$CDM and $\chi^2_{\rm min}=1739.28$ for the VIDVM model. The information criteria also prefer IDVM, with $\Delta{\rm AIC}=-9.95$ and $\Delta{\rm BIC}=-4.42$ relative to $\Lambda$CDM, indicating that the improvement in fit compensates for the additional interaction parameter. This preference is further supported by the Bayesian evidence, which yields a Bayes factor of $\Delta\log Z=+1.82$ relative to $\Lambda$CDM, corresponding to weak evidence in favor of the interacting scenario.
In contrast, the viscous extension is generally disfavored by the statistical criteria. While the AIC penalty remains relatively modest with $\Delta{\rm AIC}=+2.16$, the BIC and Bayesian evidence provide a stronger preference against the model, yielding $\Delta{\rm BIC}=+13.23$ and $\Delta\log Z=-2.43$, respectively.

\indent After including the late-time structure growth measurements the statistical preference for the interacting scenario weakens. Although the IDVM model still produces a slightly lower chi-square value, $\chi^2_{\rm min}=1748.77$, compared to $\chi^2_{\rm min}=1752.41$ for $\Lambda$CDM, the information criteria no longer strongly support the interacting extension. In particular, the IDVM model yields $\Delta{\rm AIC}=-1.64$, which indicates marginal improvement over the standard cosmological model, while the Bayesian Information Criterion shifts toward disfavouring the interaction sector with $\Delta{\rm BIC}=+3.90$. The Bayesian evidence also decreases relative to $\Lambda$CDM, producing $\Delta \log Z=-2.27$. This indicates that the inclusion of growth measurements introduces additional tension within the strongly interacting branch. The VIDVM model exhibits intermediate behaviour, yielding $\chi^2_{\rm min}=1749.31$ with $\Delta{\rm AIC}=+0.91$, $\Delta{\rm BIC}=+11.98$, and $\Delta \log Z=-2.68$. While the viscous extension modestly improves the structure growth sector, the statistical penalty associated with the additional parameters outweighs the improvement in the total fit.

\indent For the full BASE+LSS+CMB dataset combination, the statistical estimators strongly drive all extended cosmological scenarios toward the $\Lambda$CDM limit. The minimum chi-square values for all three models become nearly degenerate, with $\chi^2_{\rm min}=1762.23$ for $\Lambda$CDM, $\chi^2_{\rm min}=1761.93$ for IDVM, and $\chi^2_{\rm min}=1760.66$ for VIDVM. Consequently, the differences in the reduced chi-square values become negligible, all remaining close to $\chi^2_\nu \approx 0.937$. However, once the parameter penalties are incorporated through the information criteria, the extended models become statistically disfavored. The interacting model yields $\Delta{\rm AIC}=+1.71$ and $\Delta{\rm BIC}=+7.24$, while the viscous extension gives $\Delta{\rm AIC}=+2.46$ and $\Delta{\rm BIC}=+13.52$. The Bayesian evidence similarly favors the standard cosmological model, with $\Delta \log Z=-4.43$ for IDVM and $\Delta \log Z=-6.64$ for VIDVM.

Overall, the statistical analysis reveals that the interacting dark vacuum scenario provides some improvement over $\Lambda$CDM when only late-time background datasets are considered, primarily due to its ability to accommodate the local $H_0$ measurement. However, the inclusion of structure growth measurements weakens this preference, while the addition of CMB distance priors strongly suppresses deviations from the standard cosmological evolution. The viscous extension is statistically competitive in terms of AIC, but becomes disfavored once the Bayesian criteria are taken into account.

\begin{figure*}
\centering
\includegraphics[width=0.85\textwidth]{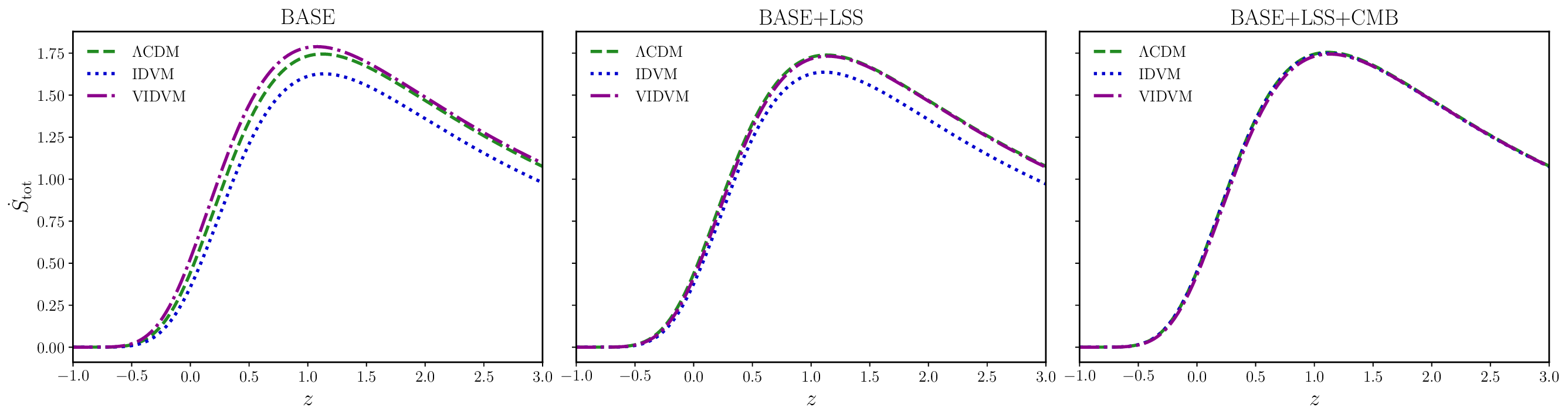}
\caption{Evolution of the total entropy production rate $\dot{S}_{\rm tot}$ for the $\Lambda$CDM, IDVM, and VIDVM cosmologies using the best-fit model parameters obtained from the three dataset combinations. The positivity of $\dot{S}_{\rm tot}$ throughout the cosmic evolution confirms the validity of the Generalized Second Law (GSL) of thermodynamics for all models considered in this work. The region $z<0$ corresponds to the future evolution of the Universe.}\label{fig9}
\end{figure*}

\begin{figure*}
\centering
\includegraphics[width=0.85\textwidth]{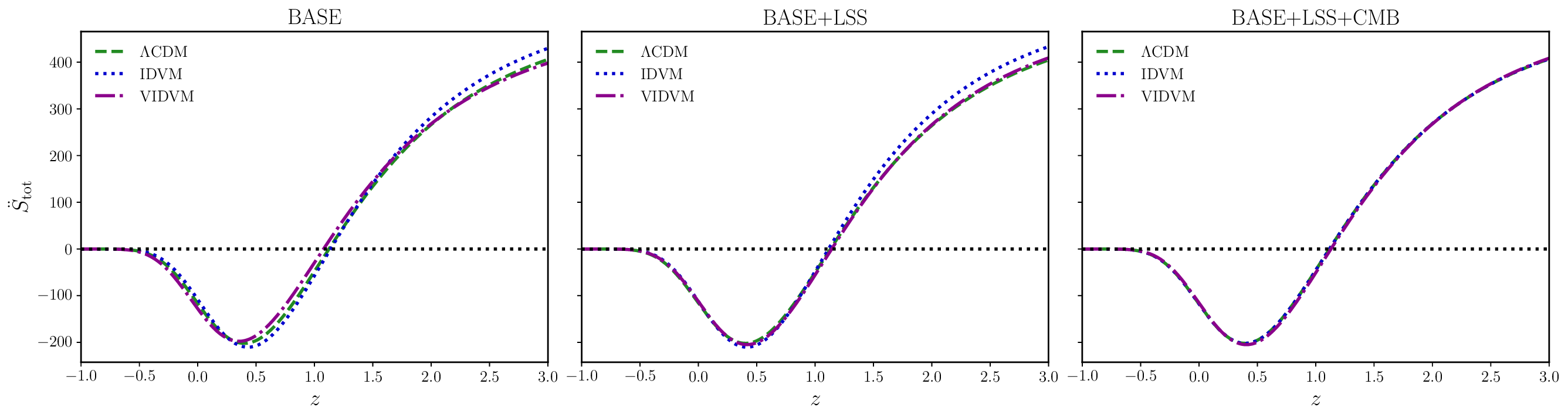}
\caption{Evolution of the second derivative of the total entropy $\ddot{S}_{\rm tot}$ for the $\Lambda$CDM, IDVM, and VIDVM scenarios. The approach of $\ddot{S}_{\rm tot}$ toward negative values at late times indicates that the Universe evolves toward a state of thermodynamic equilibrium. The future evolution region ($z<0$) further demonstrates the asymptotic thermodynamic stability of the cosmological solutions considered in this analysis.}\label{fig10}
\end{figure*}

\section{Thermodynamic Analysis}\label{sec7}

We now investigate the thermodynamic behavior of the extended viscous scenario (VIDVM) and examine the validity of the Generalized Second Law (GSL) of thermodynamics \cite{cai05}. The GSL requires that the total entropy of the Universe, including the entropy of the horizon and the cosmic fluid enclosed within it, must satisfy
\begin{equation}\label{eq46}
\dot{S}_{\rm tot} \geq 0,
\end{equation}
or equivalently,
\begin{equation}\label{eq47}
\frac{d}{dt}\left(S_H + S_M + S_\Lambda \right)\geq0.
\end{equation}

Following the Bekenstein--Hawking formalism \cite{beke73,hawk75}, the entropy associated with the cosmological horizon is given by
\begin{equation}\label{eq48}
S_H=\frac{k_B A}{4l_p^2},
\end{equation}
where $k_B$ denotes the Boltzmann constant and $l_p$ is the Planck length. The area of the Hubble horizon is $A=4\pi R_H^2$ with the horizon radius defined as
\begin{equation}\label{eq49}
R_H=\frac{1}{\sqrt{H^2+ka^{-2}}}.
\end{equation}
For a spatially flat Universe ($k=0$), this reduces to  $R_H=H^{-1}$. Assuming reduced Planck units $k_B=\hbar=c=8\pi G=1$, the horizon entropy becomes
\begin{equation}\label{eq50}
S_H=\frac{8\pi^2}{H^2},
\end{equation}
and its time derivative is
\begin{equation}\label{eq51}
\dot{S}_H=-\frac{16\pi^2\dot{H}}{H^3}.
\end{equation}

To determine the entropy of the cosmic fluid enclosed within the horizon, we employ the Gibbs equation
\begin{equation}\label{eq52}
TdS=d(\rho V)+pdV,
\end{equation}
where the volume enclosed by the Hubble horizon is
\begin{equation}\label{eq53}
V=\frac{4\pi}{3H^3},
\end{equation}
and the associated Bekenstein--Hawking temperature \cite{cai05} is
\begin{equation}\label{eq54}
T=\frac{H}{2\pi}.
\end{equation}

Since the vacuum component satisfies $p_\Lambda=-\rho_\Lambda$, it does not contribute explicitly to the Gibbs relation for the cosmic fluid entropy. Thus, the entropy of the effective cosmic fluid may be written as
\begin{equation}\label{eq55}
S_{\rm cf}=\frac{(\rho_m+p_m)V}{T}.
\end{equation}

\noindent Including both baryonic matter and viscous dark matter, we obtain
\begin{equation}\label{eq56}
S_{\rm cf}=
\frac{
(\rho_b+\rho_{dm}+p_b+\bar{p}_{dm})V
}{T},
\end{equation}
where $p_b=0$, and the effective viscous pressure is given by $\bar{p}_{dm}=-3\zeta H$. Using Eq. \eqref{eq8}, the viscous pressure reduces to $\bar{p}_{dm}=-\xi_0\rho_{dm}^{1/2}$.\\

\noindent Hence,
\begin{equation}\label{eq57}
S_{\rm cf}=
\frac{
(\rho_b+\rho_{dm}-\xi_0\rho_{dm}^{1/2})V
}{T}.
\end{equation}

Dividing by the critical density $\rho_{c,0}$ and expressing the result in terms of the density parameters yields
\begin{equation}\label{eq58}
S_{\rm cf}
=
\frac{8\pi^2}{H^2}
\left(
\frac{
\Omega_{b}a^{-3}
+
\Omega_{dm}a^{-3+\epsilon}
+
\frac{1}{2}\Omega_\zeta a^{-3+\epsilon/2}
}{E^2}
\right).
\end{equation}

Using Eq.\eqref{eq25}, this expression can be written more compactly in terms of the effective EoS parameter $w(a)$ as
\begin{equation}\label{eq59}
S_{\rm cf}
=
\frac{8\pi^2}{H^2}[1+w(a)].
\end{equation}

The total entropy is obtained by summing the horizon entropy and the cosmic fluid entropy,
\begin{equation}\label{eq60}
S_{\rm tot}
=
\frac{8\pi^2}{H^2}[2+w(a)].
\end{equation}

To compute the entropy production rate of the cosmic fluid, we differentiate the Gibbs relation with respect to cosmic time,
\begin{equation}\label{eq61}
T\dot{S}_{\rm cf}
=
(\rho+p)\dot{V}
+
\dot{\rho}V.
\end{equation}

For the complete cosmic fluid consisting of baryons and the interacting viscous dark sector, we obtain
\begin{equation}\label{eq62}
T\dot{S}_{\rm cf}
=
3H_0^2
\left(
\Omega_{b}a^{-3}
+
\Omega_{dm}a^{-3+\epsilon}
+
\frac{\Omega_\zeta}{2} a^{\frac{-3+\epsilon}{2}}
\right)
(\dot{V}-3HV).
\end{equation}

\noindent Further simplification yields
\begin{equation}\label{eq63}
\dot{S}_{\rm cf}
=
\frac{12\pi^2}{H}
[1+w(a)][1+3w(a)].
\end{equation}

\noindent The total entropy production rate then becomes
\begin{equation}\label{eq64}
\dot{S}_{\rm tot}
=
\dot{S}_H+\dot{S}_{\rm cf}.
\end{equation}

\noindent Substituting the expressions obtained in Eqs. \eqref{eq51} and \eqref{eq63}, we finally obtain
\begin{equation}\label{eq65}
\dot{S}_{\rm tot}
=
\frac{36\pi^2}{H}[1+w(a)]^2 \geq 0.
\end{equation}

\indent The positivity of $\dot S_{\rm tot}$ throughout the cosmic evolution indicates that the viscous interacting vacuum scenario remains thermodynamically viable despite the presence of dissipative effects and energy transfer in the dark sector. Thus, the extended scenario remains compatible with the GSL of thermodynamics.\\
\indent We plot the entropy evolution for all three cosmological scenarios and dataset combinations in Figs. \ref{fig9} and \ref{fig10}. Figure \ref{fig9} shows the first derivative of the total entropy, $\dot S_{\rm tot}$, which remains positive throughout the evolution, confirming the validity of the GSL. Figure \ref{fig10} displays the second derivative, $\ddot S_{\rm tot}$, which becomes negative at late times and gradually approaches zero in the asymptotic future. This behavior indicates that the Universe evolves toward a state of thermodynamic equilibrium, consistent with the expected thermodynamic fate of an accelerating cosmological spacetime \cite{radi10,pavo13}.

\section{Conclusion}\label{sec8}

\indent In this work, we investigated an interacting dark sector scenario consisting of a decaying vacuum energy density and a dark matter component having non-standard evolution. Additionally, bulk viscosity was introduced in the dark matter component to assess how non-equilibrium effects can influence a coupled dark sector. The analysis was performed using three successive observational dataset combinations. Bayesian parameter estimation was carried out using nested sampling techniques, enabling a simultaneous assessment of parameter constraints, model selection statistics, and overall viability of the cosmological scenarios. We now turn to the implications of these results for the major observational tensions in contemporary cosmology.

\indent The local measurements of the Hubble constant have remained in significant tension with values inferred from early-Universe observations. The SH0ES-R22 measurement reported $H_0=73.04\pm1.04\,{\rm km\,s^{-1}\,Mpc^{-1}}$ \cite{ries22}, while the more recent H0DN collaboration obtained the tighter constraint $H_0=73.50\pm0.80\,{\rm km\,s^{-1}\,Mpc^{-1}}$ \cite{stef26}. The persistence of this discrepancy has motivated a wide range of extensions beyond the standard cosmological model. To investigate the impact of our interacting dark sector scenario on the Hubble tension, we adopted the H0DN measurement as an independent data point in the analysis.
For the BASE dataset, the inferred values of $H_0$ correspond to tensions of $4.57\sigma$, $0.23\sigma$, and $4.83\sigma$ with the H0DN determination for the $\Lambda$CDM, IDVM, and VIDVM models respectively. Thus, the interacting vacuum model is remarkably successful in accommodating the local measurement. A similar trend persists when the LSS data are included, with the IDVM scenario exhibiting only a $0.32\sigma$ discrepancy, while both $\Lambda$CDM and VIDVM remain in approximately $4.5\sigma$ tension with the local value.
The situation changes considerably once the CMB distance priors are incorporated. The inferred Hubble constants become strongly consistent across the three cosmological scenarios, resulting in tensions of $5.23\sigma$, $4.57\sigma$, and $4.49\sigma$ with the H0DN measurement for the $\Lambda$CDM, IDVM, and VIDVM models respectively. At the same time, comparison with the Planck determination, $H_0=67.4\pm0.5\,{\rm km\,s^{-1}\,Mpc^{-1}}$, yields more moderate discrepancies of $2.99\sigma$, $2.81\sigma$, and $2.81\sigma$. The inclusion of the CMB distance priors therefore drives all three cosmological scenarios toward a common intermediate value of $H_0$, lying between the local and Planck determinations, while largely eliminating the strong late-time interacting solution favored by the BASE and BASE+LSS datasets.

\indent For structure growth, with the BASE+LSS data, the inferred $S_8$ constraints correspond to tensions of $0.72\sigma$, $1.19\sigma$, and $0.25\sigma$ with the DES weak-lensing measurement for the $\Lambda$CDM, IDVM, and VIDVM models respectively. Although the interacting vacuum model exhibits the largest departure from the preferred weak-lensing value, the discrepancy remains statistically insignificant, while the viscous extension provides marginally better agreement than the standard cosmological model. In particular, the lower $S_8$ value obtained for IDVM indicates a suppression of matter clustering. However, the negative interaction parameter contributes positively through the RSD correction, leading to an enhanced $f\sigma_8(z)$ evolution, as illustrated in Fig. \ref{fig8}. With the inclusion of the CMB distance priors, the inferred values of $S_8$ become highly consistent across all three cosmological scenarios, yielding tensions of only $0.70\sigma$, $0.66\sigma$, and $0.46\sigma$ respectively. Overall, the structure growth predictions of the cosmological scenarios remain in good agreement with current weak-lensing observations, with no statistically significant tension with the DES estimate.

\indent The dark-sector dynamics are governed by the interaction parameter $\epsilon$ and the viscous density parameter $\Omega_\zeta$, with the former playing a significant role in earlier epochs and the latter becoming prominent only in late-times. Notably, the inclusion of CMB distance priors drives $\epsilon$ toward zero, effectively suppressing the direct interaction between dark matter and vacuum energy. In contrast, $\Omega_\zeta$ remains nonzero and admits a larger value in comparison to the BASE and BASE+LSS constraints. This suggests that while dark-sector interactions are disfavored by early-Universe observations, the bulk viscous contribution associated with our choice of $\zeta$ remains compatible with current data and may still influence the late-time evolution of the dark sector.\\
\indent We also find that the transition redshift $z_t$ shifts to earlier epochs in the IDVM for the BASE and BASE+LSS datasets, while the $\Lambda$CDM and VIDVM predictions remain similar. With the inclusion of CMB priors, however, the inferred values of $z_t$ become statistically consistent across all models. The constraints on $q_0$, $z_t$, $w_0$, and $j_0$, together with the evolution of the deceleration parameter $q(z)$, effective equation-of-state parameter $w(z)$, and jerk parameter $j(z)$ shown in Figs.~\ref{fig5}--\ref{fig7}, indicate that both interacting scenarios remain kinematically consistent with the standard cosmological picture. In particular, the inclusion of CMB distance priors causes the predicted kinematic histories of the extended models to converge toward those of $\Lambda$CDM.

\indent From a statistical standpoint, the IDVM scenario is favored by the BASE dataset, with strong support from the AIC and mild preference from the BIC and Bayes factor. The inclusion of LSS information weakens this preference, although the interacting model continues to outperform $\Lambda$CDM according to the AIC. For the BIC and Bayesian evidence, however, the model becomes mildly disfavored. Once the CMB priors are incorporated, $\Lambda$CDM emerges as the preferred cosmological scenario, with weak evidence against both extended models according to the AIC and stronger evidence according to the BIC and Bayes factor. The VIDVM scenario exhibits a different behavior. Although the viscous extension does not provide a significant statistical improvement over the pure interacting model, it remains competitive with $\Lambda$CDM according to the AIC for the BASE+LSS dataset. For the remaining dataset combinations, the evidence against the viscous scenario ranges from weak to mild. The BIC and Bayesian evidence, however, consistently impose a stronger penalty on the additional model complexity, resulting in a systematic preference for the concordance cosmology. Overall, the background-only observations mildly favor the purely interacting scenario, largely due to its ability to alleviate the $H_0$ tension. The inclusion of progressively stronger observational constraints reduces this preference and drives the extended models toward the standard cosmological picture.\\
\indent Thermodynamically, both the interacting scenario and the viscous extension remain viable, with the generalized second law of thermodynamics remaining satisfied throughout the recent past and future evolution of the Universe. Nevertheless, the present framework contains several phenomenological assumptions. The interaction parameter $\epsilon$ modifies the standard dark matter scaling, while the bulk viscous coefficient $\zeta$ is introduced through a phenomenological ansatz that approximates a constant bulk viscosity in the matter dominated regime. Although these ingredients provide a consistent realization of the decaying vacuum scenario, their underlying origin remains unexplored.
A further limitation arises from the use of the Eckart formalism which is non-causal. A more rigorous treatment would require a causal description such as the Israel--Stewart framework \cite{isra79}, which introduces relaxation effects and an explicit time dependence in the evolution of the bulk viscous pressure. In addition, our perturbation analysis incorporates only the effects of the dark sector interaction and does not explicitly include viscous corrections at the perturbative level. While this approximation is justified by the small values obtained for $\Omega_\zeta$, typically of order $10^{-3}$, a fully self-consistent treatment of viscous perturbations remains an important direction for future investigation.\\
\indent In summary, the modified dark-sector scenarios considered in this work provide a consistent framework for exploring departures from the standard cosmological model. While the interacting vacuum sector is capable of alleviating the late-time $H_0$ tension and remains compatible with current structure-growth observations, the inclusion of early-Universe information strongly suppresses the interaction strength and drives the cosmological evolution toward the $\Lambda$CDM limit. In contrast, the bulk viscous contribution remains comparatively resilient to these constraints, surviving as a small but potentially relevant late-time modification of the dark sector. Although the statistical preference for the extended models weakens with increasingly stringent datasets, the persistence of the viscous component, together with its thermodynamic viability, suggests that dissipative effects may provide an interesting avenue for exploring alternatives to $\Lambda$CDM. Future investigations employing causal viscous frameworks and fully consistent perturbation treatments may further clarify the role of bulk viscosity and energy exchange in a unified dark sector cosmology.\\\\

\acknowledgments
L.C. would like to thank University Grant Commission (UGC), India for providing Junior Research Fellowship (JRF) to carry out this work.


%
%


\begin{thebibliography}{99}
%
\bibitem{agha20} N. Aghanim \textit{et al.} (Planck Collaboration), \emph{Planck 2018 results-V. CMB power spectra and likelihoods}, Astron. Astrophys. \textbf{641} (2020) A5
\bibitem{bern02} F. Bernardeau, S. Colombi, E. Gazta\~naga, and R. Scoccimarro, \emph{Large-scale structure of the Universe and cosmological perturbation theory}, Phys. Rep. \textbf{367} (2002) 1
\bibitem{ries98} A. G. Riess, A. V. Filippenko, P. Challis, A. Clocchiatti, A. Diercks, P. M. Garnavich, R. L. Gilliland, C. J. Hogan, S. Jha, R. P. Kirshner, \textit{et al.}, \emph{Observational Evidence from Supernovae for an Accelerating Universe and a Cosmological Constant}, Astron. J. \textbf{116} (1998) 1009
\bibitem{perl99} S. Perlmutter, G. Aldering, G. Goldhaber, R. A. Knop, P. Nugent, P. G. Castro, S. Deustua, S. Fabbro, A. Goobar, D. E. Groom, \textit{et al.} (The Supernova Cosmology Project), \emph{Measurements of $\Omega$ and $\Lambda$ from 42 High-Redshift Supernovae}, Astrophys. J. \textbf{517} (1999) 565
\bibitem{wein89} S. Weinberg, \emph{The cosmological constant problem}, Rev. Mod. Phys. \textbf{61} (1989) 1
\bibitem{carr01} S. M. Carroll, \emph{The cosmological constant}, Living Rev. Relativ. \textbf{4} (2001) 1
\bibitem{peeb03} P. J. E. Peebles and B. Ratra, \emph{The cosmological constant and dark energy}, Rev. Mod. Phys. \textbf{75} (2003) 559
\bibitem{cope06} E. J. Copeland, M. Sami, and S. Tsujikawa, \emph{Dynamics of dark energy}, Int. J. Mod. Phys. D \textbf{15} (2006) 1753
\bibitem{kari25} M. A. Karim \textit{et al.} (DESI Collaboration), \emph{DESI DR2 results. II. Measurements of baryon acoustic oscillations and cosmological constraints}, Phys. Rev. D \textbf{112} (2025) 083515
\bibitem{abbo24} T. M. C. Abbott \textit{et al.} (DES Collaboration), \emph{The Dark Energy Survey: Cosmology Results with $\sim1500$ New High-redshift Type Ia Supernovae Using the Full 5 yr Data Set}, Astrophys. J. Lett. \textbf{973} (2024) L14
\bibitem{tsuj13} S. Tsujikawa, \emph{Quintessence: a review}, Class. Quantum Grav. \textbf{30} (2013) 214003
\bibitem{sola18a} J. Sol\'{a} , \emph{Running vacuum in the universe: Current phenomenological status}, in \textit{Proceedings of The Fourteenth Marcel Grossmann Meeting on General Relativity}, edited by M. Bianchi, R. T. Jantzen, and R. Ruffini (World Scientific, Singapore, 2017), pp. 2363--2370.
\bibitem{zeld68} Y. B. Zeldovich, \emph{The cosmological constant and the theory of elementary particles}, Sov. Phys. Usp. \textbf{11} (1968) 381
\bibitem{ozer86} M. Ozer and M. O. Taha, \emph{A possible solution to the main cosmological problems}, Phys. Lett. B \textbf{171} (1986) 363
\bibitem{free87} K. Freese, F. C. Adams, J. A. Frieman, and E. Mottola, \emph{Cosmology with decaying vacuum energy}, Nucl. Phys. B \textbf{287} (1987) 797
\bibitem{shap03} I. L. Shapiro and J. Sol\`a, \emph{The scaling evolution of the cosmological constant}, J. High Energy Phys. \textbf{02} (2002) 006
\bibitem{borg05} H. A. Borges and S. Carneiro, \emph{Friedmann cosmology with decaying vacuum density}, Gen. Relativ. Gravit. \textbf{37} (2005) 1385
\bibitem{sola13} J. Sol\`a, \emph{Cosmological constant and vacuum energy: Old and new ideas}, J. Phys. Conf. Ser. \textbf{453} (2013) 012015
\bibitem{sola15} J. Sol\`a, A. G\'omez-Valent, and J. de Cruz P\'erez, \emph{Hints of dynamical vacuum energy in the expanding Universe}, Astrophys. J. Lett. \textbf{811} (2015) L14
\bibitem{sola18b} J. Sol\`a Peracaula, A. G\'omez-Valent, and J. de Cruz P\'erez, \emph{The $H_0$ tension in light of vacuum dynamics in the Universe}, Phys. Lett. B \textbf{774} (2017) 317
\bibitem{zhan19} J.-J. Zhang, C.-C. Lee, and C.-Q. Geng, \emph{Observational constraints on running vacuum model}, Chin. Phys. C \textbf{43} (2019) 025102
\bibitem{sing21} C. P. Singh and J. Sol\`a Peracaula, \emph{Friedmann cosmology with decaying vacuum density in Brans--Dicke theory}, Eur. Phys. J. C \textbf{81} (2021) 960
\bibitem{kaeo23} C. Kaeonikhom, H. Assadullahi, J. Schewtschenko, and D. Wands, \emph{Observational constraints on interacting vacuum energy with linear interactions}, J. Cosmol. Astropart. Phys. \textbf{01} (2023) 042
\bibitem{sola23u} J. Sol\`a Peracaula, A. G\'omez-Valent, J. de Cruz P\'erez, and C. Moreno-Pulido, \emph{Running Vacuum in the Universe: Phenomenological Status in Light of the Latest Observations, and Its Impact on the $\sigma_8$ and $H_0$ Tensions}, Universe \textbf{9} (2023) 262
\bibitem{brit25} L. S. Brito, J. F. Jesus, A. A. Escobal, and S. H. Pereira, \emph{Can decaying vacuum solve the $H_0$ tension?}, Eur. Phys. J. C \textbf{85} (2025) 1025
\bibitem{chan26a} L. Chander and C. P. Singh, \emph{Gravitationally-induced matter creation cosmology with power-law decaying vacuum energy}, Eur. Phys. J. Plus \textbf{141} (2026) 33
\bibitem{chan26b} L. Chander and C. P. Singh, \emph{Matter Creation Cosmology With Decaying Vacuum Energy: Observational and Thermodynamical Analyses}, Fortschr. Phys. \textbf{74} (2026) e70105
\bibitem{cruz26} J. de Cruz P\'erez, A. G\'omez-Valent, and J. Sol\`a Peracaula, \emph{Dynamical dark energy models in light of the latest observations}, Phys. Rev. D \textbf{113} (2026) 083521
\bibitem{hu98} W. Hu, \emph{Structure Formation with Generalized Dark Matter}, Astrophys. J. \textbf{506} (1998) 485
\bibitem{piat11} O. F. Piattella, J. C. Fabris, and W. Zimdahl, \emph{Bulk viscous cosmology with causal transport theory}, J. Cosmol. Astropart. Phys. \textbf{05} (2011) 029
\bibitem{velt11} H. Velten and D. J. Schwarz, \emph{Constraints on dissipative unified dark matter}, J. Cosmol. Astropart. Phys. \textbf{09} (2011) 016
\bibitem{ecka40} C. Eckart, \emph{The Thermodynamics of Irreversible Processes. III. Relativistic Theory of the Simple Fluid}, Phys. Rev. \textbf{58} (1940) 919
\bibitem{land87} L. D. Landau and E. M. Lifshitz, \textit{Fluid Mechanics}, 2nd ed. (Pergamon, Oxford, 2013).
\bibitem{maar96} R. Maartens, \emph{Causal Thermodynamics in Relativity}, arXiv:astro-ph/9609119.
\bibitem{barro86} J. D. Barrow, \emph{The deflationary universe: An instability of the de Sitter universe}, Phys. Lett. B \textbf{180} (1986) 335
\bibitem{zimd96} W. Zimdahl, \emph{Bulk viscous cosmology}, Phys. Rev. D \textbf{53} (1996) 5483
\bibitem{brev05} I. Brevik and O. Gorbunova, \emph{Dark energy and viscous cosmology}, Gen. Relativ. Gravit. \textbf{37} (2005) 2039
\bibitem{fab06} J. C. Fabris, S. V. B. Gonalves, and R. de Sa Ribeiro, \emph{Bulk viscosity driving the acceleration of the Universe}, Gen. Relativ. Gravit. \textbf{38} (2006) 495
\bibitem{sing07} C. P. Singh, S. Kumar, and A. Pradhan, \emph{Early viscous universe with variable gravitational and cosmological constants}, Class. Quantum Grav. \textbf{24} (2007) 455
\bibitem{sing08} C. P. Singh, \emph{Bulk viscous cosmology in early Universe}, Pramana J. Phy.\textbf{71} (2008) 33
\bibitem{sing18} C.P. Singh and A. Kumar, \emph{Ricci dark energy model with bulk viscosity}, Eur. Phys. J. Plus {\bf133} (2018) 312
\bibitem{sing18a} C.P. Singh and M. Srivastava, \emph{Viscous cosmology in new holographic dark energy model and the cosmic acceleration}, Eur. Phys. J. C {\bf78} (2018) 190
\bibitem{simr23} S. Kaur and C.P. Singh, \emph{Viscous cosmology in holographic dark energy with Granda-Oliveros cut-off}, Commun. Theor. Phys. {\bf75} (2023) 025401
\bibitem{avel09} A. Avelino and U. Nucamendi, \emph{Can a matter-dominated model with constant bulk viscosity drive the accelerated expansion of the universe?}, J. Cosmol. Astropart. Phys. \textbf{04} (2009) 006
\bibitem{norm16} B. D. Normann and I. Brevik, \emph{General Bulk-Viscous Solutions and Estimates of Bulk Viscosity in the Cosmic Fluid}, Entropy \textbf{18}, 215 (2016).
\bibitem{barb17} C. M. S. Barbosa, H. Velten, J. C. Fabris, and R. O. Ramos, \emph{Assessing the impact of bulk and shear viscosities on large-scale structure formation}, Phys. Rev. D \textbf{96} (2017) 023527
\bibitem{most18} B. Mostaghel, H. Moshafi, and S. M. S. Movahed, \emph{The integrated Sachsolfe effect in the bulk viscous dark energy model}, Mon. Not. R. Astron. Soc. \textbf{481} (2018) 1799
\bibitem{avel25} P. P. Avelino, A. R. Gomes, and D. A. Tamayo, \emph{Note on bulk viscosity as an alternative to dark energy}, Phys. Rev. D \textbf{112} (2025) 123531
\bibitem{isra79} W. Israel and J. M. Stewart, \emph{Transient relativistic thermodynamics and kinetic theory}, Ann. Phys. \textbf{118} (1979) 341
\bibitem{sara21} N. Sarath, N. D. Jerin Mohan, and T. K. Mathew, \emph{Running vacuum cosmology with bulk viscous matter}, Mod. Phys. Lett. A \textbf{38} (2023) 2350099
\bibitem{cruz23} N. Cruz, G. G\'omez, E. Gonz\'alez, G. Palma, and A. Rinc\'on, \emph{Exploring models of running vacuum energy with viscous dark matter from a dynamical system perspective}, Phys. Dark Universe \textbf{42} (2023) 101351
\bibitem{sing24} C. P. Singh and V. Khatri, \emph{Viscous fluid dynamics with decaying vacuum energy density}, Phys. Rev. D \textbf{109} (2024) 023508
\bibitem{nand24} T. Nandi and A. Choudhuri, \emph{Bulk viscous matter interacting with decaying vacuum energy density: A model for late-time evolution of the Universe}, Eur. Phys. J. C \textbf{85} (2025) 782
\bibitem{khat25a} V. Khatri, C.P. Singh and M. Srivastava, \emph{Exploring interacting bulk viscous model with deacying vacuum density}, Astr. Compt. {\bf53} (2025) 100992
\bibitem{khat25} V. Khatri and C. P. Singh, \emph{Interacting model of bulk viscous and decaying vacuum energy}, Phys. Lett. B \textbf{871} (2025) 139994
\bibitem{ma95} C.-P. Ma and E. Bertschinger, \emph{Cosmological Perturbation Theory in the Synchronous and Conformal Newtonian Gauges}, Astrophys. J. \textbf{455} (1995) 7
\bibitem{perc10} W. J. Percival, B. A. Reid, D. J. Eisenstein, N. A. Bahcall, T. Budav\'ari, J. A. Frieman, M. Fukugita, J. E. Gunn, \v{Z}. Ivezi\'c, G. R. Knapp, \textit{et al.}, \emph{Baryon acoustic oscillations in the Sloan Digital Sky Survey Data Release 7 galaxy sample}, Mon. Not. R. Astron. Soc. \textbf{401} (2010) 2148
\bibitem{basi14} S. Basilakos and J. Sol\`a, \emph{ Growth index of matter perturbations in running vacuum models}, Phys. Rev. D \textbf{92} (2015) 123501
\bibitem{gome17} A. G\'omez-Valent and J. Sol\`a, \emph{Vacuum models with a linear and a quadratic term in $H$: Structure formation and number counts analysis}, Mon. Not. R. Astron. Soc. \textbf{448} (2015) 2810
\bibitem{spea20} J. S. Speagle, \emph{DYNESTY: a dynamic nested sampling package for estimating Bayesian posteriors and evidences}, Mon. Not. R. Astron. Soc. \textbf{493}, 3132 (2020).
\bibitem{skil04} J. Skilling, \emph{Nested sampling}, AIP Conf. Proc. \textbf{735} (2004) 395
\bibitem{skil06} J. Skilling, \emph{Nested sampling for general Bayesian computation}, Bayesian Anal. \textbf{1} (2006) 833
\bibitem{buch17} J. Buchner, \emph{A statistical test for Nested Sampling algorithms}, Stat. Comput. \textbf{26} (2016) 383
\bibitem{more20} M. Moresco, L. Amati, L. Amendola, S. Birrer, J. P. Blakeslee, M. Cantiello, A. Cimatti, J. Darling, M. Della Valle, M. Fishbach, \textit{et al.}, \emph{Unveiling the Universe with emerging cosmological probes}, Living Rev. Relativ. \textbf{25} (2022) 6
\bibitem{stef26} S. Casertano \textit{et al.} (H0DN Collaboration), \emph{The Local Distance Network: A community consensus report on the measurement of the Hubble constant at $\sim1\%$ precision}, Astron. Astrophys. \textbf{708} (2026) A166
\bibitem{alam21} S. Alam, M. Aubert, S. Avila, C. Balland, J. E. Bautista, M. A. Bershady, D. Bizyaev, M. R. Blanton, A. S. Bolton, \textit{et al.}, \emph{Completed SDSS-IV extended Baryon Oscillation Spectroscopic Survey: Cosmological implications from two decades of spectroscopic surveys at the Apache Point Observatory}, Phys. Rev. D \textbf{103} (2021) 083533
\bibitem{abbo25} T. M. C. Abbott \textit{et al.} (DES Collaboration), \emph{Dark energy survey year 3 results: Cosmological constraints from cluster abundances, weak lensing, and galaxy clustering}, Phys. Rev. D \textbf{112} (2025) 083535
\bibitem{aka73} H. Akaike, \emph{A new look at the statistical model identification}, IEEE Trans. Autom. Control \textbf{19} (1974) 716
\bibitem{sch78} G. Schwarz, \emph{Estimating the Dimension of a Model}, Ann. Statist. \textbf{6} (1978) 461
\bibitem{kass95} R. E. Kass and A. E. Raftery, \emph{Bayes factors}, J. Am. Stat. Assoc. \textbf{90} (1995) 773
\bibitem{beke73} J. D. Bekenstein, \emph{Black Holes and Entropy}, Phys. Rev. D \textbf{7} (1973) 2333
\bibitem{hawk75} S. W. Hawking, \emph{Particle creation by black holes}, Commun. Math. Phys. \textbf{43} (1975) 199
\bibitem{gab23} G. G\'{o}mez, G. Palma, E. Gonz\'{a}lez, A. Rinc\'{o}n and N. Cruz, A new parametrization for bulk viscosity cosmology as extension of the $\Lambda$CDM
model, Eur. Phys. J. C {\bf138} (2023) 738
\bibitem{eeh17} E. Mostaghel, H. Moshafi and S.M.S. Movahed, Non-minimal derivative coupling scalar field and bulk viscous dark energy, Eur. Phys. J. C {\bf77} (2017) 541
\bibitem{shap09} I. L. Shapiro and J. Sol\`a, \emph{On the possible running of the cosmological ``constant''}, Phys. Lett. B \textbf{682} (2009) 105
\bibitem{sola17} J. Sol\`a, A. G\'omez-Valent, and J. de Cruz P\'erez, \emph{First evidence of running cosmic vacuum: Challenging the concordance model}, Astrophys.  J. \textbf{836} (2017) 43
\bibitem{sola22} J. Sol\`a Peracaula, \emph{The cosmological constant problem and running vacuum in the expanding Universe}, Phil. Trans. R. Soc. A \textbf{380} (2022) 20210182
\bibitem{basi09} S. Basilakos, M. Plionis, and J. Sol\`a, Hubble expansion and structure formation in time varying vacuum models, Phys. Rev. D \textbf{80} (2009) 083511
\bibitem{alca05} J. S. Alcaniz and J. A. S. Lima, Interpreting cosmological vacuum decay, Phys. Rev. D \textbf{72} (2005) 063516
\bibitem{cost10} F. E. M. Costa and J. S. Alcaniz, Cosmological consequences of a possible $\Lambda$-dark matter interaction, Phys. Rev. D \textbf{81} (2010) 043506
\bibitem{wang16} B. Wang, E. Abdalla, F. Atrio-Barandela, and D. Pav\'{o}n, Dark matter and dark energy interactions: theoretical challenges, cosmological implications and observational signatures, Rep. Prog. Phys. \textbf{79} (2016) 096901
\bibitem{gome18} A. G\'{o}mez-Valent and J. Sol\`{a}  Peracaula, \emph{Density perturbations for running vacuum: a successful approach to structure formation and to the $\sigma_8$-tension}, Mon. Not. R. Astron. Soc. \textbf{478} (2018) 126
\bibitem{silv21} W. J. C. da Silva and R. Silva, \emph{Growth of matter perturbations in the extended viscous dark energy models}, Eur. Phys. J. C \textbf{81} (2021) 403
\bibitem{yang18} W. Yang, A. Mukherjee, E. Di Valentino, and S. Pan, \emph{Interacting dark energy with time varying equation of state and the $H_0$ tension}, Phys. Rev. D \textbf{98} (2018) 123527
\bibitem{trip98} M. Tripp, \emph{A two-parameter luminosity correction for Type Ia supernovae}, Astron. Astrophys. \textbf{331} (1998) 815
\bibitem{conl11} A. Conley, J. Guy, M. Sullivan, N. Regnault, P. Astier, C. Balland, S. Basa, R. G. Carlberg, D. Fouchez, D. Hardin, \textit{et al.}, \emph{Supernova Constraints and Systematic Uncertainties from the First Three Years of the Supernova Legacy Survey}, Astrophys. J. Suppl. Ser. \textbf{192} (2011) 1
\bibitem{scol18} D. M. Scolnic, D. O. Jones, A. Rest, Y. C. Pan, R. Chornock, R. J. Foley, M. E. Huber, R. Kessler, G. Narayan, A. G. Riess, et al., \emph{The Complete Light-curve Sample of Spectroscopically Confirmed SNe Ia from Pan-STARRS1 and Cosmological Constraints from the Combined Pantheon Sample}, Astrophys. J. \textbf{859} (2018) 101
\bibitem{eise05} D. J. Eisenstein, I. Zehavi, D. W. Hogg, R. Scoccimarro, M. R. Blanton, R. C. Nichol, R. Scranton, H.-J. Seo, M. Tegmark, Z. Zheng, \textit{et al.},\emph{ Detection of the Baryon Acoustic Peak in the Large-Scale Correlation Function of SDSS Luminous Red Galaxies}, Astrophys. J. \textbf{633} (2005) 560
\bibitem{alam17} S. Alam, M. Ata, S. Bailey, F. Beutler, D. Bizyaev, J. A. Blazek, A. S. Bolton, J. R. Brownstein, A. Burden, C.-H. Chuang, \textit{et al.}, \emph{The clustering of galaxies in the completed SDSS-III Baryon Oscillation Spectroscopic Survey: cosmological analysis of the DR12 galaxy sample}, Mon. Not. R. Astron. Soc. \textbf{470} (2017) 2617
\bibitem{jime02} R. Jimenez and A. Loeb, \emph{Constraining Cosmological Parameters Based on Relative Galaxy Ages}, Astrophys. J. \textbf{573} (2002) 37
\bibitem{more12} M. Moresco, A. Cimatti, R. Jimenez, L. Pozzetti, G. Zamorani, M. Bolzonella, J. Dunlop, F. Lamareille, M. Mignoli, H. Pearce, \textit{et al.}, \emph{Improved constraints on the expansion rate of the Universe up to $z\sim1.1$ from the spectroscopic evolution of cosmic chronometers}, J. Cosmol. Astropart. Phys. \textbf{08} (2012) 006
\bibitem{more16} M. Moresco, L. Pozzetti, A. Cimatti, R. Jimenez, C. Maraston, L. Verde, D. Thomas, A. Citro, R. Tojeiro, and D. Wilkinson, \emph{A 6\% measurement of the Hubble parameter at $z\sim0.45$: direct evidence of the epoch of cosmic re-acceleration}, J. Cosmol. Astropart. Phys. \textbf{05} (2016) 014
\bibitem{more18} M. Moresco, R. Jimenez, L. Verde, L. Pozzetti, A. Cimatti, and A. Citro, \emph{Setting the Stage for Cosmic Chronometers. I. Assessing the Impact of Young Stellar Populations on Hubble Parameter Measurements}, Astrophys. J. \textbf{868} (2018) 84
\bibitem{chen19} L. Chen, Q.-G. Huang, and K. Wang, \emph{Distance priors from Planck final release}, J. Cosmol. Astropart. Phys. \textbf{02} (2019) 028
\bibitem{wang07} Y. Wang and P. Mukherjee, \emph{Observational constraints on dark energy and cosmic curvature}, Phys. Rev. D \textbf{76} (2007) 103533
\bibitem{cybu16} R. H. Cyburt, B. D. Fields, K. A. Olive, and T.-H. Yeh, \emph{Big bang nucleosynthesis: Present status}, Rev. Mod. Phys. \textbf{88} (2016) 015004
\bibitem{trot08} R. Trotta, \emph{Bayes in the sky: Bayesian inference and model selection in cosmology}, Contemp. Phys. \textbf{49} (2008) 71
\bibitem{cai05} R.-G. Cai and S. P. Kim, \emph{First law of thermodynamics and Friedmann equations of Friedmann-Robertson-Walker universe}, J. High Energy Phys. \textbf{02} (2005) 050
\bibitem{radi10} N. Radicella and D. Pav\'{o}n, \emph{A thermodynamic motivation for dark energy}, Gen. Relativ. Gravit. \textbf{44} (2012) 685
\bibitem{pavo13} D. Pav\'{o}n and N. Radicella, \emph{Does the entropy of the Universe tend to a maximum?}, Gen. Relativ. Gravit. \textbf{45} (2013) 63
\bibitem{ries22} A. G. Riess, W. Yuan, L. M. Macri, D. Scolnic, D. Brout, S. Casertano, D. O. Jones, Y. Murakami, G. S. Anand, L. Breuval, \textit{et al.}, \emph{A Comprehensive Measurement of the Local Value of the Hubble Constant with 1 km s$^{-1}$ Mpc$^{-1}$ Uncertainty from the Hubble Space Telescope and the SH0ES Team}, Astrophys. J. Lett. \textbf{934} (2022) L7


\end{thebibliography}
\end{document}